\documentclass[aps,prd,twocolumn,floatfix,nofootinbib]{revtex4}
\usepackage{graphicx} \usepackage{epsf}
 
\def\gsim{ \lower .75ex \hbox{$\sim$} \llap{\raise .27ex \hbox{$>$}} }
\def\lsim{ \lower .75ex \hbox{$\sim$} \llap{\raise .27ex \hbox{$<$}} }
\def\be{\begin{equation}}  \def\ee{\end{equation}}  
      
  \def\mpl{M_{\rm pl}}

  \def\mpl{M_{Pl}}  
 
\begin{document} 
 
\title{Observing Trans-Planckian Signatures in the Cosmic Microwave
Background}
 
\author{Richard Easther$^{1}$}  \author{William H. Kinney$^{2}$}
\author{Hiranya Peiris$^3$\footnote{Hubble Fellow}}
 
\affiliation{~} \affiliation{$^1$Department of Physics, Yale
University, New Haven  CT 06520, USA} \affiliation{$^2$Dept.\ of
Physics, University at Buffalo, SUNY, Buffalo, NY 14260, USA \\ Email:
{\tt whkinney@buffalo.edu}} \affiliation{$^3$Department of
Astrophysical Sciences, Princeton University, Princeton NJ 08544, USA
and\\ Kavli Institute for Cosmological Physics,Enrico Fermi Institute,
University of Chicago, Chicago IL 60637, USA \\ Email: {\tt
hiranya@cfcp.uchicago.edu}}
 
\begin{abstract} 
We examine the constraints cosmological observations can place on any
trans-Planckian corrections to the primordial spectrum of
perturbations underlying the anisotropies in the Cosmic Microwave
Background.  We focus on models of trans-Planckian physics which lead
to a modulated primordial spectrum.  Rather than looking at a generic
modulated spectrum, our calculations are based on a specific model,
and are intended as a case study for the sort of constraints one could
hope to apply on a well-motivated model of trans-Planckian physics.
We present results for three different approaches -- a grid search in
a subset of the overall parameter space, a Fisher matrix estimate of
the likely error ellipses, and a Monte Carlo Markov Chain fit to a
simulated CMB sky.  As was seen in previous analyses,  the likelihood
space has multiple peaks, and we show that their distribution  can be
reproduced via a simple semi-analytic argument. All three methods lead
to broadly similar results.   We vary 10 cosmological parameters
(including two related to the trans-Planckian terms) and show that the
amplitude of the tensor perturbations is directly correlated with the
detectability of any trans-Planckian modulation. We argue that this
is likely to be true for any trans-Planckian modulation in the
paradigm of slow-roll inflation.   For the specific case we consider,
we conclude that  if the tensor to scalar ratio, $r \sim 0.15$, the
ratio between the inflationary Hubble scale $H$, and the scale of new
physics $M$ has to be on the order of $0.004$ if the modulation is
detectable at the 2$\sigma$ level.  For a lower value of $r$, the
bound on $H/M$ becomes looser.
\end{abstract} 
 
 \maketitle

\section{Introduction}
 
Conventional cosmology is based upon a four dimensional spacetime
whose evolution is governed by the Einstein field equations.   This
picture certainly breaks down at the Planck scale  due to quantum
corrections to the gravitational action. Furthermore,  stringy effects
may change our understanding of spacetime at energies one or two
orders of magnitude below the Planck scale. Unfortunately, both the
Planck and string energy scales are entirely inaccessible to direct
experimentation.  However, thanks to the dramatic expansion of the
universe during the inflationary era, it is possible that cosmology
provides an unexpected window into physics at the shortest {\em
length\/} scales \cite{Brandenberger:1999sw}-\cite{Shankaranarayanan:2004iq}.   

If we assume the universe has always been matter dominated and is  now
around 13 billion years old, the current horizon size is roughly 40
billion lightyears, or $2\times 10^{61}$ Planck lengths.  If inflation
ended at the GUT scale ($\sim 10^{15}$GeV) with efficient reheating,
the universe has grown roughly  $4\times10^{27}$ times larger since
reheating, and by at least a similar amount during inflation, in order
to ensure that the universe is homogeneous on scales equal to the
present horizon size.  Consequently, all distances significantly
smaller than galactic scales in the current universe were necessarily
sub-Planckian at some point during inflation.   If the inflationary
period continues for longer than the minimum time necessary to solve
the cosmological flatness and horizon problems, all presently visible
scales must have crossed the Planckian boundary during inflation.
Conversely, without inflation  all astrophysical scales are mapped to
lengths larger than the Planck length at all times since the age of the 
universe was of order the Planck time.

On the face of it, the fact that astrophysical scales in the present
universe correspond to sub-Planckian scales during the inflationary
epoch is simply a mathematical curiosity. However,   inflation
generates perturbations through the exponential  stretching of quantum
fluctuations.   The textbook calculation of the primordial
perturbation spectrum  implicitly ignores the consequences of the breakdown of spacetime at some new fundamental lengthscale. For a clear exposition of the standard treatment of perturbations, including a discussion of the effect of boundary conditions, see \cite{Liddlebk}.  In string theory the effective minimum length can be
significantly larger than the Planck length, by perhaps one or two
orders of magnitude
\cite{Kaplunovsky:1987rp,Arkani-Hamed:1998rs,Antoniadis:1998ig}.  In
this case, the ratio between the inflationary Hubble scale, $H$, and
the scale of new stringy physics, $M$ can be as large as $10^{-2}$.

Dimensional analysis alone suggests that any modifications to the
perturbation spectrum induced by including this minimum length will be
a function of $(H/M)^p$, where $p$ is, of course, unknown.  If $p$ is
unity, then there is at least a corner of parameter space where we
might expect  potentially observable modifications  to the
spectrum. However, for larger values of $p$, any  modifications to the
spectrum are most likely unobservable. String theory and other
candidate unified models are not sufficiently mature to permit an {\em
ab initio\/} analysis of perturbation generation, and  predict the
value of $p$. Consequently, all studies of trans-Planckian
modifications to the inflationary perturbation spectrum  require some
ansatz which, while motivated by our understanding of stringy /
Planckian physics, introduces a degree of arbitrariness into the
calculation.  This ansatz necessarily modifies either the initial
conditions or the evolution of the perturbations.   Generically,
modifications to the initial conditions produce changes in the
spectrum on the order of $H/M$,\footnote{See, however, Ref.
 \cite{Bozza:2003pr} for a discussion of this issue in terms of
minimization of the Hamiltonian on an initial hypersurface.} whereas
modifications to the evolution result in changes on the order of
$(H/M)^2$. It is worth noting, however, that no calculation has
predicted that the modifications vanish completely.

Normally, this lack of theoretical consensus would deter detailed
studies of the observational consequences of any given model of
trans-Planckian  physics. However, even a small possibility that the
fingerprints of Planckian or stringy physics can be found on the sky
provides  more than enough motivation for us to take a careful look at
what constraints can  be established with present data or future
measurements.     In this paper, we focus on one model of
trans-Planckian physics, first suggested by Danielsson in the case of
de Sitter space \cite{Danielsson:2002kx} and generalized to  non-de
Sitter backgrounds by Easther, Greene, Kinney and Shiu
\cite{Easther:2002xe}.  There is nothing intrinsically special about
this model -- our analysis is effectively a case study of what might
be achieved when one has a specific model to test. In particular, we
do not enter into the debate as to whether or not such ``truncated
$\alpha$-vacua'' are self-consistent with respect to being
well-defined in the ultraviolet limit or with respect to their
suitability as initial conditions for inflation. We simply consider
the model as a reasonable case study for investigating the
discriminatory power of forthcoming CMB observations.

The one general statement we can make is that it is far more likely
that  trans-Planckian effects  will have an observable impact on the
spectrum if inflation takes place at a relatively high energy scale.
In principle, inflation can occur at almost any scale high enough to
allow for baryogenesis and nucleosynthesis in the post-inflationary
universe -- and this can be far below the GUT scale.   However, since
we only expect a detectable signal if $H/M$ is comparatively large, we
are pushed into the corner of parameter space where $M$ is small
(compared to the Planck scale) while $H$ is roughly GUT-scale.   A
small $M$  (two orders of magnitude below the Planck  scale, say) is
entirely consistent with string theory, whereas $H$ is effectively a
free parameter.  An upper bound on $H$ arises from the limits on the
tensor component to the primordial perturbation spectrum -- which is
potentially observable via both the temperature anisotropies in the
CMB, and the B-mode of the CMB polarization. (A rough upper limit on
$H$ from the WMAP data is $H < 6 \times 10^{14}\ {\rm GeV}$.)  It is
very easy for inflation to happen at a low enough scale to ensure that
a primordial tensor spectrum is effectively unobservable by any
conceivable experiment \cite{Lyth:1996im,Knox:2002pe,Kesden:2002ku,Seljak:2003pn}.
However, if we were to observe a contribution to the CMB that appears
to be related to Planckian or stringy physics, there is a strong
likelihood that we would be in the region of parameter space where $H$
was large enough in order to ensure that the tensor contribution was
also detectable.

There are important consequences to this qualitative correlation
between  trans-Planckian contributions to the CMB and the tensor
perturbations.   First, it further raises the stakes for high
precision measurements of the CMB. We already know that an observable
tensor spectrum fixes the inflationary energy scale and rules out some
of inflation's competitors, such as the ekpyrotic \cite{Khoury:2001wf}
and pre-Big Bang scenarios \cite{Gasperini:1992em,Lidsey:1999mc}, and
that single field models of inflation must satisfy a set of
consistency conditions \cite{Lidsey:1995np}.  However, on top of these
achievements, determining $H$ would let us put direct constraints on
the scale of new physics in any specific model of trans-Planckian
physics, since $M$ would be the only remaining free parameter. If $H$
turns out to be at the high end of the permitted range, it increases
the likelihood that we will be able to  see any trans-Planckian signal
that is there to be found.  Finally, one us [RE] recently argued that
a high value of $H$ during inflation is naturally correlated with a
scalar spectrum with a significant running index ($d n_s/d\ln{k} \ne 0$)
\cite{Easther:2004ir}. We do not consider models with substantially
broken scale invariance in this paper, but do incorporate a non-zero
running into our analysis below.

From a practical perspective, if we perform a fit to CMB data with a
model that includes trans-Planckian effects, it is important to
include the tensor modes in any analysis, since the same parameter
($H$) that governs their amplitude and scale dependence is also likely
to appear in the trans-Planckian corrections.  Conversely, simply
observing an oscillatory spectrum would not be enough to prove that
Planck / string scale physics had left an imprint on the CMB -- one
can  produce the same result purely with physics at lower energy 
\cite{Burgess:2002ub,Sriramkumar:2004pj}.
 
This problem has been addressed in several previous analyses. The
first  computation  of trans-Planckian modifications  to the spectrum
was \cite{Bergstrom:2002yd}, which used a Fisher matrix calculation to
predict that the Planck mission will put stringent bounds on a
oscillatory component in the fundamental spectrum.   However, this
analysis marginalized  over a small subset of cosmological parameters,
and the use of a Fisher matrix in this context is problematic
\cite{Elgaroy:2003gq}. Subsequent calculations have been less
optimistic.   In particular, \cite{Elgaroy:2003gq} argued that any
reasonable modulation to the spectrum is most likely unobservable,
although we will argue that their pessimism is probably misplaced.
Okamoto and Lim \cite{Okamoto:2003wk} looked at the constraints on a
general superimposed oscillation to the spectrum. They varied several
cosmological parameters as well as those describing the oscillation,
and concluded that values of $H/M \gsim 0.005$ might be reached with a
cosmic variance limited survey, but ignored the role of tensor
modes. Martin and Ringeval,
\cite{Martin:2003sg,Martin:2004iv,Martin:2004yi} have studied  models
with independent parameters governing the amplitude and frequency of
the trans-Planckian modulation.  Their analysis is distinguished by
the possibility that very high-frequency oscillations with
non-negligible amplitude are considered, with the frequency of the
oscillations producing the primary constraint on $H/M$. In such  a
case, they conclude that very strong constraints are possible on the
value of $H/M$, even when restricted to  {\em current} data.

This paper is organized as follows: In Sec. \ref{sec:model}, we
discuss the specific model of trans-Planckian physics we consider in
this paper.  The modulation to the power spectrum is correlated to the
tensor/scalar ratio, which we argue is well motivated because of its
relationship to a $k$-independent  boundary condition. In
Sec. \ref{sec:quicksurveys}, we discuss two complementary  surveys of
the trans-Planckian parameter space: First, we consider a grid-based
search of the parameter space, which is computationally inefficient
but enables a broad sweep of the space. Second, we consider a
Fisher-matrix based analysis, which handles large numbers of
parameters in an efficient way, but does not sample the global
properties of the likelihood surface. In Sec.  \ref{sec:markovchains},
we describe a Markov Chain Monte Carlo analysis, which constitutes a
complementary method for exploring the parameter space.  All three
analyses are in good agreement. Comments and conclusions are presented
in Sec. \ref{sec:conclusions}.
 
\section{The Model}
\label{sec:model}

Quantum fluctuations during inflation produce two fluctuation spectra:
 the curvature perturbation in the comoving gauge  ${\cal R}$, and the
 two polarization states of the primordial tensor perturbation, $h_+$
 and $h_\times$.  We parameterize these power spectra by
\begin{eqnarray}
 \label{eq:P_R} \Delta^2_{\cal R}(k)&=& \Delta^2_{\cal R}(k_\star)
  \left(\frac{k}{k_\star}\right)^{n_s(k_\star)-1+\frac{1}{2}(dn_s/d\ln
  k)\ln(k/k_\star)}, \\ \label{eq:P_h} \Delta^2_h(k)&=&
  \Delta^2_h(k_\star)
  \left(\frac{k}{k_\star}\right)^{n_t(k_\star)+\frac{1}{2}(dn_t/d\ln
  k)\ln(k/k_\star)},
\end{eqnarray}
where $\Delta^2(k_\star)$ is a normalization constant, and $k_\star$
is some pivot wavenumber.  The running, $dn/d\ln k$, is defined by the
second derivative of the  power spectrum, $dn/d\ln k\equiv
d^2\Delta^2/d\ln k^2$, for both the  scalar and the tensor modes.
This parameterization gives the definition of the spectral index,
\be
\label{eq:nsdef}
 n_s(k)-1 \equiv \frac{d \ln \Delta^2_{\cal R}}{d \ln k}
 n_s(k_\star)-1+\frac{dn_s}{d\ln k}\ln\left(\frac{k}{k_\star}\right),
 \ee
for the scalar modes, and
\be
\label{eq:ntdef}
 n_t(k) \equiv \frac{d \ln \Delta^2_h}{d \ln k} =
 n_t(k_\star)+\frac{dn_t}{d\ln k}\ln\left(\frac{k}{k_\star}\right), \ee
for the tensor modes.  In addition, we re-parameterize the tensor
power spectrum amplitude, $\Delta^2_h(k_\star)$, by the
``tensor/scalar ratio $r$'', the relative amplitude  of the
tensor-to-scalar modes, given by
\begin{equation}
 \label{eq:rdef} r \equiv \frac{\Delta^2_h(k_\star)}{\Delta^2_{\cal R}(k_\star)}.
\end{equation}
For a single slowly rolling inflaton, $n_t=-r/8$, so we can reduce the
number of parameters by fixing $n_t$ once $r$ is chosen. We choose the
pivot-scale $k_\star=0.002$ Mpc$^{-1}$ at which to evaluate $n_s$, $r$
and $A$, where $A(k_\star)$ and $\Delta^2_{\cal R}(k_\star)$ are
related \citep[see][]{Verde:2003ey} through
\begin{eqnarray}
 \label{eq:Adef} \Delta^2_{\cal R}(k_\star) &\simeq& 2.95\times10^{-9} A(k_\star).
\end{eqnarray}
Note that in this definition, the trans-Planckian parameter
$\epsilon$, which appears in Eq. \ref{Eq:TPPS}, is simply the first
slow-roll parameter \citep[see e.g.][]{Liddlebk},
\begin{equation}
\label{eq:eps}
\epsilon \equiv {m_{\rm Pl}^2 \over 4 \pi} \left({H'\left(\phi\right)
\over H\left(\phi\right)}\right) \simeq
\frac{\mpl^2}{2}\left(\frac{V'}{V}\right)^2,
\end{equation}
where $\mpl\equiv (8\pi G)^{-1/2} = m_{\rm pl}/\sqrt{8\pi}= 2.4\times
10^{18}~{\rm GeV}$ is the reduced Planck energy; it is related to the
tensor-to-scalar ratio by $r=16\epsilon$. The definition in terms of
$H(\phi)$, called the {\em Hubble slow roll} expansion, is exact, and
the expression in terms of the inflationary potential $V(\phi)$,
called the {\em potential slow roll} expansion, is an approximation
valid in the limit of a slowly rolling field, $\dot\phi^2 \ll
V(\phi)$. The primordial power spectra $\Delta^2_{\cal R}$ and
$\Delta^2_h$ are the underlying spectra which are modulated by effects
from Planck-scale physics, described below.

The specific model of trans-Planckian physics we are working with
\cite{Danielsson:2002kx,Easther:2002xe} assumes that rather than
reducing to the Minkowski  mode function $u_k$ in the infinite past,
the perturbations are matched at some finite time, corresponding to
the moment (different for each comoving perturbation) when their
physical wavelength exceeds the minimum length, $1/M$.  Following
\cite{Easther:2002xe}, we can compute the modified spectrum;
\begin{equation}
\Delta_{{\cal R},h} = \sqrt{ | C_{+} + C_{-}|}  \Delta^{BD}_{{\cal R},h}
\end{equation}
where $BD$ refers to the spectrum obtained with the conventional Bunch-Davies
vacuum.   In this case the scalar and tensor spectra are both modified
by the same amount.  The $C_{\pm}$ are   \cite{Easther:2002xe}
\begin{eqnarray}
C_{+} &=& \frac{1}{2}  \exp{\left(- \frac{i y_c}{1- \epsilon}\right)}
  \left(\frac{ 2 y_c + i}{y_c} \right) u_k(y_c) \, ,\\ C_{-} &=&-
  \frac{1}{2}  \exp{\left( \frac{i y_c}{1- \epsilon}\right)}
  \left(\frac{  i}{y_c} \right) u_k(y_c)   \, .
\end{eqnarray}
Here $y$ is a rescaled ``time'', defined by $y=k/(aH)$
\cite{Kinney:1997ne},  so that in the absence of a new scale $y$
decreases from $+\infty$ to $0$, with horizon crossing taking place
(by definition) when $y=1$.  In this language, $y_c(k)$ is the
critical time for the $k$-th mode, at which its physical scale becomes
larger than the minimum length.  This is a $k$-dependent quantity,
since in all slow roll models, $H$ is a (slowly varying) function of
time, and in general \cite{Easther:2002xe}, $y_c \propto
k^{\epsilon}$ where $\epsilon$ is the usual slow roll parameter.  This
relationship is exact when $\epsilon$ is expressed in the Hubble slow
roll expansion, and approximate when the potential slow roll form is
used. Consequently,
\be y_c(k)   = \frac{M}{H_\star}   \left(  \frac{k}{k_\star}
\right)^{\epsilon}   \ee
where $M$ is the scale of new physics, and the starred quantities
again refer to some fiducial value of $k$. The quantity $u_k(y_c)$ is
the value of the quantum mode function (given in the short-wavelength
limit by $u_k \sim e^{-i k \tau}$) evaluated when the wavelength of
the mode is equal to a fixed physical cutoff length $k / a = M$, or
equivalently $y = y_c(k)$. See Ref. \cite{Easther:2002xe} for a
detailed discussion.

Keeping only the lowest order term in $1/y_c $, we can write the
modified spectrum in a more transparent form,
\begin{equation}
\label{Eq:TPPS}
\Delta_{{\cal R},h} = \left[   1 +  \frac{1}{y_c}  \sin\left({\frac{2
y_c}{1-\epsilon} + \phi}\right)\right]^{1/4}    \Delta^{BD}_{{\cal R},h},
\end{equation}
where $\phi$ is a phase factor quantifying our lack of {\sl a priori}
knowledge about the physical ``pivot-scale'' $k_\star$.

We note that the functional form of the modification above is
determined not by the details of the boundary condition, but by the
behavior of the {\em background}, in particular the changing ratio of
the inflationary horizon size to the (fixed) fundamental scale. Such a
sinusoidal modulation likely to be a generic feature of any
trans-Planckian  ansatz that changes the initial conditions for a
given perturbation, so long as the boundary condition itself has no
intrinsic dependence on $k$.\footnote{See Ref. \cite{Ashoorioon:2004vm} 
for a discussion of the effect of boundary terms in the action on this 
prescription. An interesting alternative
prescription is discussed in Ref. \cite{Armendariz-Picon:2003gd}. }
Since this effect depends on $H$, which gradually decreases during the
course of the inflationary epoch, the original spectrum is multiplied
by a (small) modulation, with amplitude of order $H/M$  -- and, as a
consequence, the amplitude of the modulation decreases with increasing
$k$. A plot for the specific case of power-law inflation is to be
found in \cite{Easther:2002xe}, and in Fig. \ref{fig:modulation}.

\section{Quick Surveys}
\label{sec:quicksurveys}

There are several ways to estimate parameters, and their likely
uncertainties, in CMB data. In addition to the Markov Chain Monte
Carlo (MCMC) methods discussed below, two widely used approaches have
been grid searches in parameter space and Fisher matrices, which
provide theoretical estimates of the likely accuracy that can be
obtained with a given measurement.  A key observation that applies to
all methods is that we cannot simply estimate trans-Planckian
parameters after the other cosmological parameters are determined. If
we add new parameters (from trans-Planckian physics or elsewhere) we
must  re-estimate all the other parameters in our model.  Failing to
do so will result in unphysically small error ellipses.

Grid methods have no intrinsic theoretical shortcomings, but with more
than three or four variables they become far too time consuming for a
thorough search, since the number of points on the grid increases
exponentially with the number of parameters.  Fisher matrix techniques
scale well with the total number of parameters, but they cannot cope
with a likelihood surface that contains multiple peaks, and we will
see that this is precisely the situation we are confronting here. That
said, we use both methods to make a rough survey of the parameter
space before discussing our Markov chain results.

In all cases, we use a modified version of the code {\tt CMBFAST}
 4.5.1 with the high-precision option \cite{Seljak:1996is} to
 calculate the power spectra. The $\ell$--space and $k$--space
 resolution of {\tt CMBFAST} was significantly increased in order to
 resolve the trans-Planckian modulations with sufficient accuracy. We
 do not, however, calculate every $\ell$--mode, as suggested in
 \cite{Martin:2004iv} --- this would render the computational cost of
 exploring such a high-dimensional parameter space prohibitively high,
 since the computation of a single model takes $\sim 6$ minutes of
 computing time at the resolution used, roughly an order of magnitude
 longer than needed on the same hardware with an unmodified version of
 {\tt CMBFAST}. However, unlike \cite{Martin:2004iv}, we are not
 testing modulations with independent amplitudes and frequencies; in
 our model, very high frequency oscillations either have strongly
 suppressed amplitudes or enormous tensor/scalar ratios, or
 both. Thus, while we need to sample in $\ell$-space sufficiently to
 resolve the oscillations we are looking for, sampling every
 $\ell$-mode is not a requirement for the model at hand.

\subsection{Likelihood Function for an Ideal Experiment measuring ${T, E, B}$}

In order to calculate the best possible constraints obtainable for the
trans-Planckian model above, we wish to simulate an ideal noiseless
CMB experiment with full sky coverage. Assuming CMB multipoles are
Gaussian-distributed, the likelihood function for such an experiment
is given by
\begin{equation}
{\mathcal L} \propto \prod_{\ell m} \frac{\exp \left[ -\frac12 {\bf
    D}^\dagger_{\ell m} {\bf C}^{-1} {\bf D}_{\ell m}
    \right]}{\sqrt{\det {\bf C}}}.
\end{equation}
Here, ${\bf D}_{\ell m}$ is the data vector of spherical harmonic
coefficients
\begin{equation}
{\bf D}_{lm} = \left[ a_{\ell m}^T, a_{\ell m}^E, a_{\ell m}^B \right],
\end{equation}
where, for example, the temperature map has been expanded in spherical
harmonics as $\hat{T}(\hat{n}) = \sum_{\ell m} a_{\ell m} Y_{\ell m}$,
and {\bf C} is the covariance matrix given by
\begin{equation}
{\bf C} =   \left( \begin{array}{ccc}  C_\ell^{TT} & C_\ell^{TE} & 0
\\ C_\ell^{TE} & C_\ell^{EE} & 0 \\ 0 & 0 & C_\ell^{BB}
\end{array} \right).
\end{equation}
The terms $C_\ell^{TB}$ and $C_\ell^{EB}$ are zero by global
isotropy. In the covariance matrix, $C_\ell^{X Y}$ denote theoretical
power spectra. Now we define the estimator for the power spectra from
the data as:
\begin{equation}
\hat{C}_\ell^{XY} = \sum_m \frac{|a_{\ell m}^{X\dagger} a_{\ell
m}^{Y}|}{2\ell+1}.
\end{equation}
Since the universe is assumed to be isotropic, the likelihood function
is independent of $m$, and summing over it, one obtains, up to an
irrelevant constant:
\begin{eqnarray}
-2 \ln {\mathcal L} &=& \sum_\ell (2\ell+1) \left\{
 \ln\left(\frac{C_\ell^{BB}}{\hat{C}_\ell^{BB}}\right)
 \right.\nonumber \\ &+& \ln\left( \frac{C_\ell^{TT} C_\ell^{EE} -
 \left(C_\ell^{TE} \right)^2}{ \hat{C}_\ell^{TT} \hat{C}_\ell^{EE} -
 \left(\hat{C}_\ell^{TE}\right)^2} \right) \nonumber \\ &+&
 \frac{\hat{C}_\ell^{TT} C_\ell^{EE} + C_\ell^{TT} \hat{C}_\ell^{EE} -
 2 \hat{C}_\ell^{TE} C_\ell^{TE}}{C_\ell^{TT} C_\ell^{EE} -
 \left(C_\ell^{TE} \right)^2} \nonumber \\ &+&
 \left. \frac{\hat{C}_\ell^{BB}}{C_\ell^{BB}} - 3 \right\}.
\end{eqnarray}
The likelihood has been normalized with respect to the maximum
likelihood, where $C_\ell^{X Y} = \hat{C}_\ell^{X Y}$, so that it
behaves like a $\chi^2$ statistic.

Given a set of observational data points, the goal is then to compute
the  likelihood function for various choices of parameters, {\it i.e.}
a Bayesian analysis. Since we are interested in forecasting the
performance of future measurements, the ``data'' is synthetic rather
than from an actual observation.  Various methods for performing the
likelihood analysis have  advantages and disadvantages with respect to
computational efficiency and coverage of the parameter space. In the
next section, we discuss the results of a grid-based search of the
trans-Planckian parameter space.

\subsection{Grids}
\label{sec:grid}

Grid calculations are a simple ``brute force'' method of exploring a
given parameter space. Given a set of observational data points (for
example a CMB multipole spectrum), the Bayesian best-fit $N$-parameter
model is found by dividing the $N$-dimensional parameter space into a
grid of evenly spaced points and computing the likelihood function of
each model relative to the data. Error bars are assigned by the rate
of dropoff of the likelihood function from the best-fit point, which
in a coarse grid typically involves interpolation between the
likelihoods actually computed on the grid.. Since the number of
calculations increases exponentially with the number of parameters
$N$, grid methods are inefficient for exploring large parameter spaces.

\begin{figure}
\includegraphics[width=3.1in]{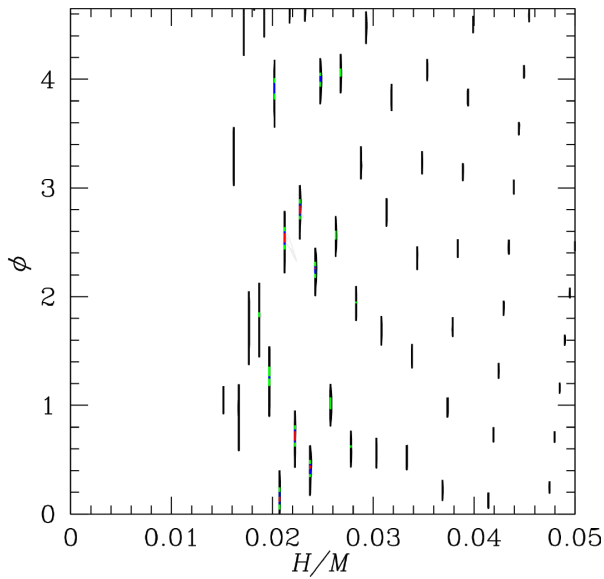}
\caption{\label{fig:coarsegrid} A coarse grid (100 points in the range
$H/M = [0,0.05]$ by 75 points in the range $\phi = [0,3 \pi /2]$)
covering a broad region of the trans-Planckian parameter space. The
inner (colored) contours are drawn at the $1\sigma$, $2\sigma$, and
$3\sigma$ levels, and the outer (black) contours are at a
$\Delta\chi^2 = 50$ relative to the best fit to better show the shape
of the likelihood function. The fiducial model has $H/M= 0.022$ and
$\phi = 0.021$. }
\end{figure}

\begin{figure}
\includegraphics[width=3.1in]{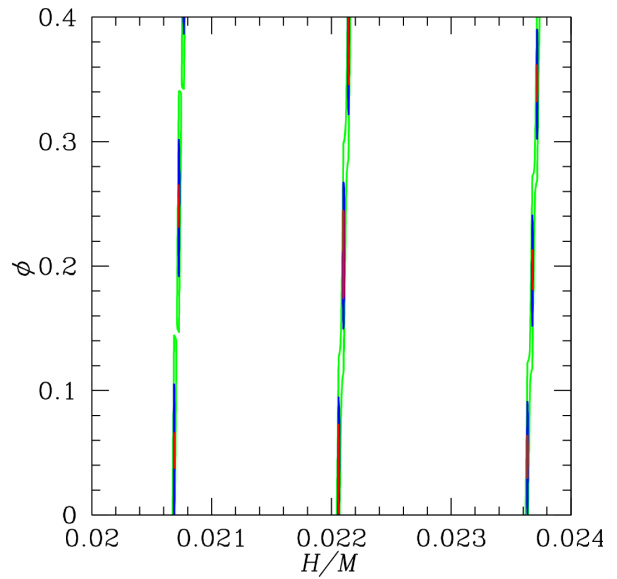}
\caption{\label{fig:finegrid} A fine grid (100 points in the range
$H/M = [0.020,0.024]$ by 100 points in the range $\phi = [0,0.4]$)
showing detail in a small region of the parameter space from
Fig. \ref{fig:coarsegrid}. The fiducial model is in the center of the
plot, and contours are drawn at $1\sigma$ (inner, red), $2\sigma$
(middle, blue), and $3\sigma$ (outer, green) levels. }
\end{figure}

Since we are interested in forecasting the capabilities of future CMB
measurements we do not use current  observational data points,  but
produce ``synthetic'' data by computing a fiducial model representing
the assumed underlying cosmology and assign a set of error bars to the
points in the fiducial model representing the accuracy of the
observation. Our hypothetical baseline experiment  is cosmic variance
limited to $\ell = 1500$ in all four CMB spectra: $C^{TT}_\ell$,
$C^{TE}_\ell$, $C^{EE}_\ell$, and $C^{BB}_\ell$. While this is an
unrealistic assumption, it characterizes the amount of information
which is intrinsically contained in the CMB spectra themselves, that
is: What is the best we can possibly do? Since grid methods are
particularly well suited to performing a broad survey of a small
parameter space, we choose our parameter space to be the
trans-Planckian amplitude $H/M$, and the phase $\phi$ of the
oscillations, which is in principle arbitrary. Other cosmological
parameters are held fixed. The fiducial model we assume for the grid
calculation is approximately the best-fit from the WMAP data set,
$\Omega_b = 0.05$, $\Omega_c = 0.25$, $\Omega_\Lambda =
0.7$, $h = 0.72$, $\tau_{\rm ri} = 0.12$, $n_s = 0.95$, $dn_s/d\ln k = -0.02$. We
assume a tensor/scalar ratio of $r = 0.1$, and trans-Planckian
parameters $H/M= 0.022$ and $\phi = 0.021$. (These apparently odd
choices for the trans-Planckian parameters are chosen so that the
fiducial model lies exactly on a grid point in
Fig. \ref{fig:finegrid}.) Figure \ref{fig:coarsegrid} shows the result
of a coarse grid covering a large range of parameters, showing the
substantial degeneracy in the $H/M$ - $\phi$ plane, in particular many
``islands'' in the likelihood function representing good fits between
the test and fiducial models. Figure \ref{fig:finegrid} is a finer
grid run over a smaller region of the same parameter space, showing
more detail in the shape of the likelihood function.

\begin{figure}
\includegraphics[width=3.1in]{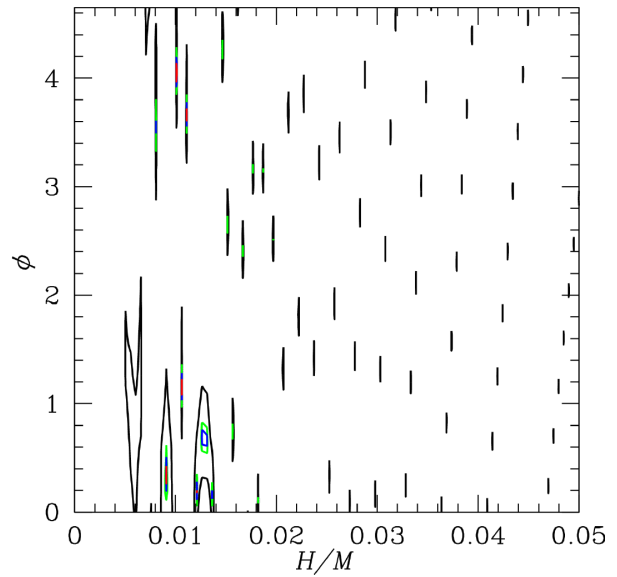}
\caption{\label{fig:smallamplitude} Grid results for a fiducial model
with $H/M = 0.01$ and $\phi = 2.0$, showing the broadening of the
degeneracy for smaller amplitude. (Note that the fiducial model is
missed due to finite grid size effects.)}
\end{figure}

\begin{figure}
\includegraphics[width=3.1in]{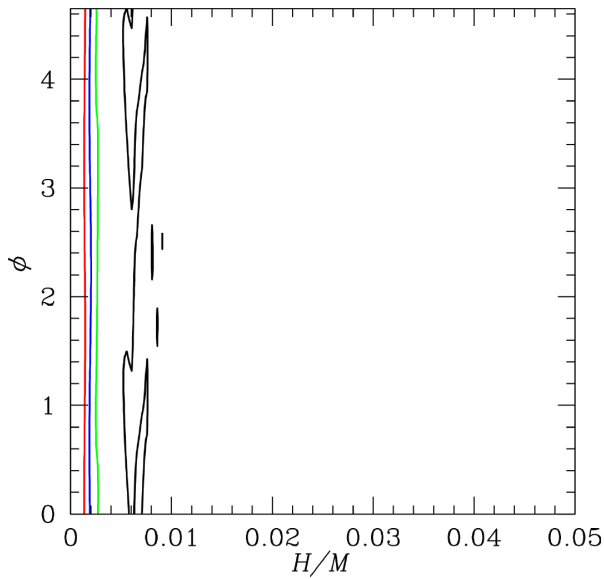}
\caption{\label{fig:null} Grid results for the null hypothesis, $H/M =
0$, estimating  an upper limit on $H/M$ if no trans-Planckian signal
is detected.}
\end{figure}

Several conclusions can be drawn from this analysis. First, the
projected error ellipses for our hypothetical cosmic-variance limited
measurement are extremely small in the $H/M$ plane, indicating a clear
ability to detect the trans-Planckian oscillations in the power
spectrum. However, because of the presence of a large number of nearly
degenerate ``islands'' in the likelihood function, it will be
difficult to pin down the precise values of both the trans-Planckian
parameters, especially the phase.  This is consistent with the results
obtained in previous analyses
\cite{Elgaroy:2003gq,Okamoto:2003wk}. However, we do note that for the
fiducial model chosen, {\em none} of the best-fit ``islands'' is
consistent with zero amplitude, $H/M = 0$. Therefore, although the
values of trans-Planckian parameters are poorly constrained, the case
of zero modulation can be ruled out to high significance. Figure
\ref{fig:smallamplitude} shows the shape of the likelihood surface
relative to a fiducial model with $H/M = 0.01$, and Fig.
\ref{fig:null} shows the likelihood relative to the null hypothesis,
$H/M = 0$, which gives a measure of the size of the {\em upper limit}
one could place on $H/M$ if no signal were detected. (We emphasize
that the $3\sigma$ upper limit of about $H/M < 0.0025$ in
Fig. \ref{fig:null} does not take into account variations in other
cosmological parameters.)

\begin{figure*} 
\begin{center}
$\begin{array}{c@{\hspace{1in}}c} \epsfxsize=2.8in
\epsffile{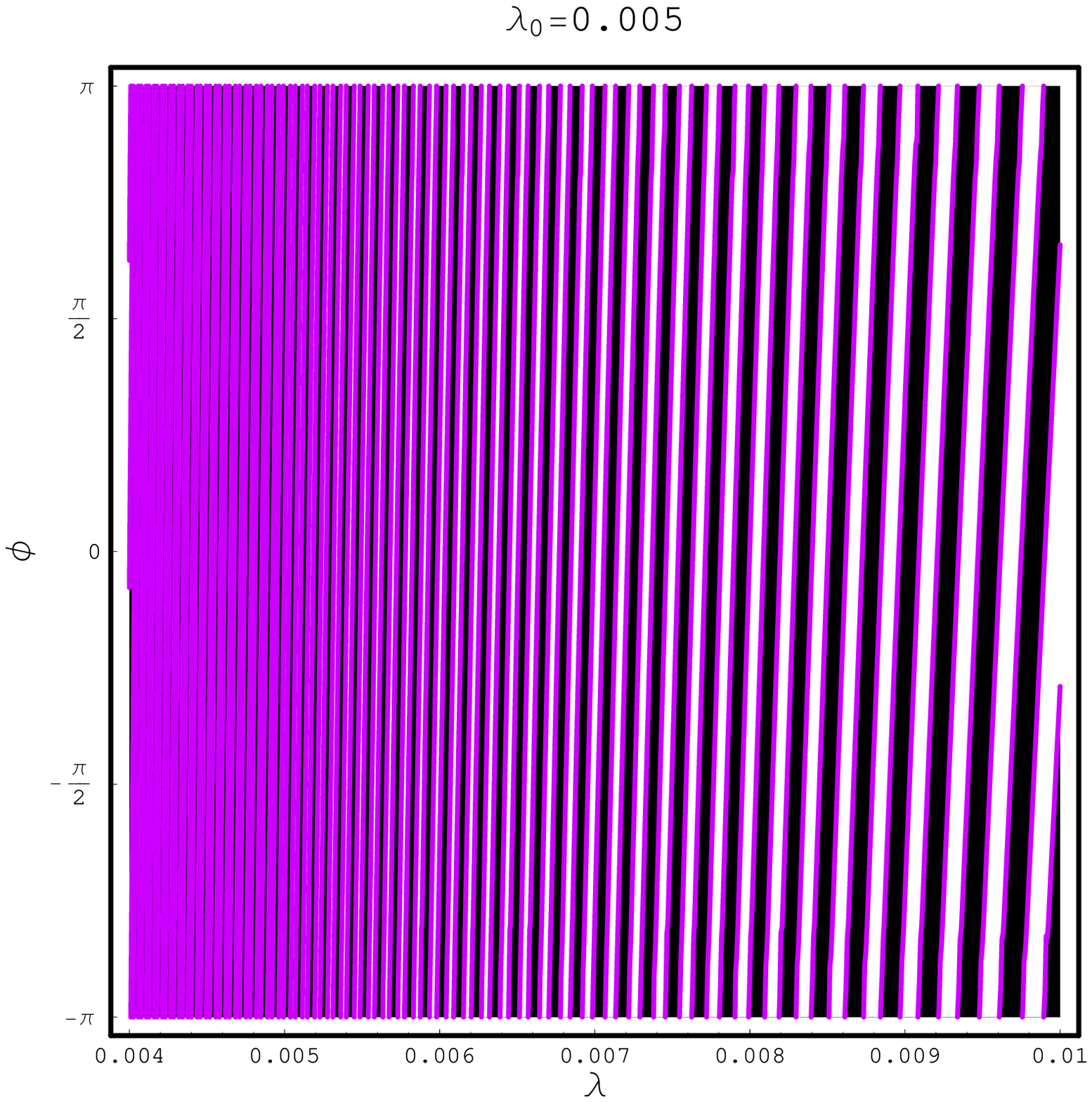} & \epsfxsize=2.8in \epsffile{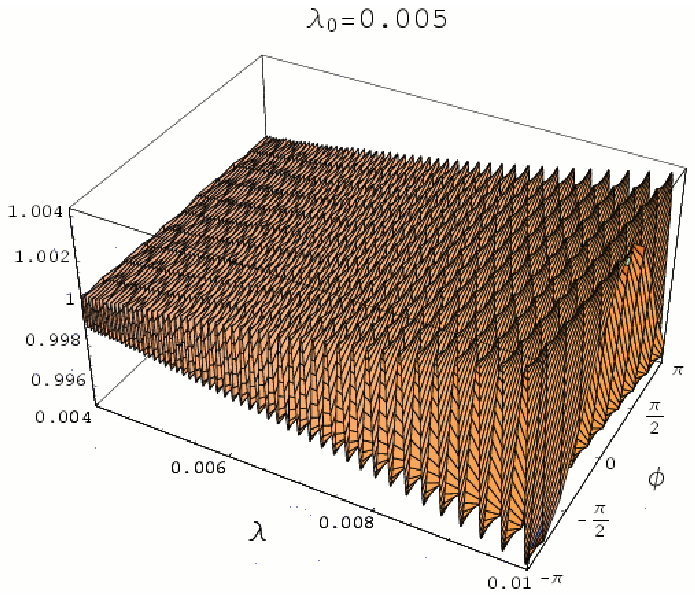} \\
\epsfxsize=2.8in \epsffile{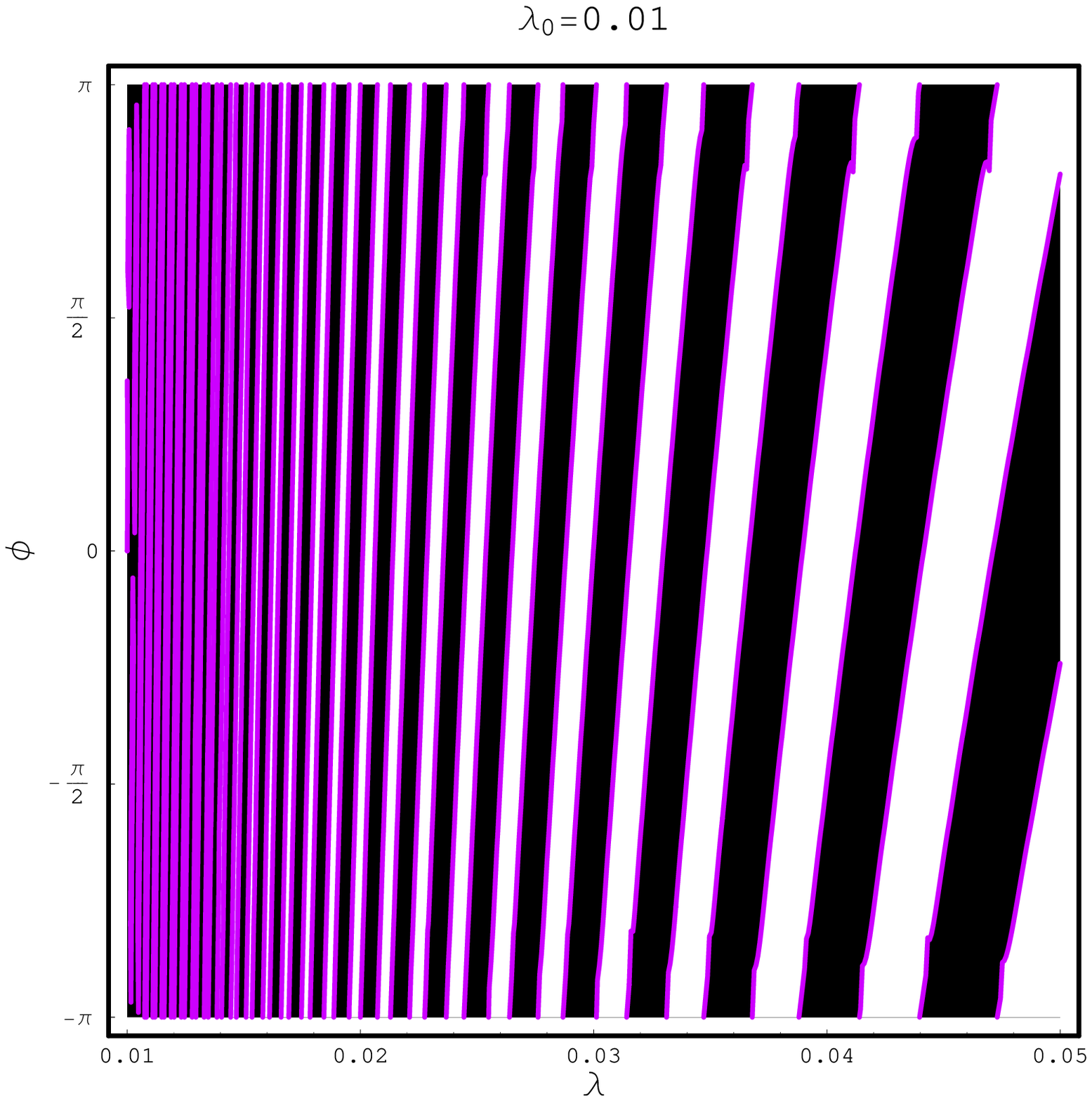} & \epsfxsize=2.8in
\epsffile{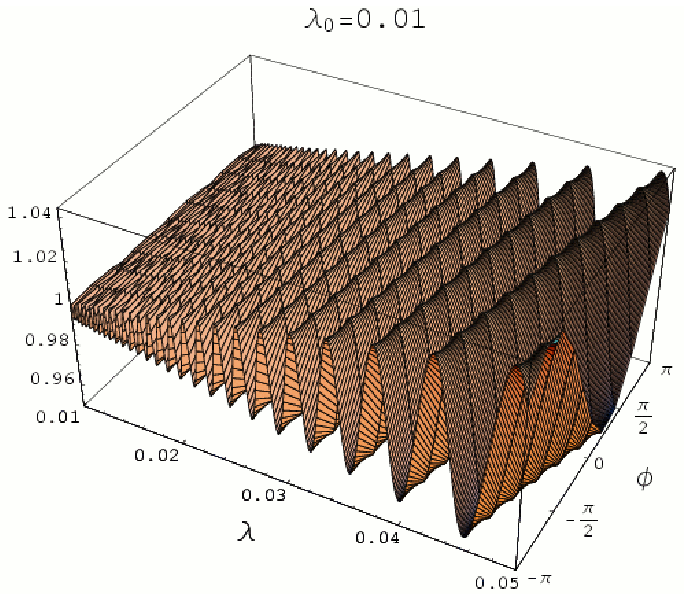} \\
\epsfxsize=2.8in \epsffile{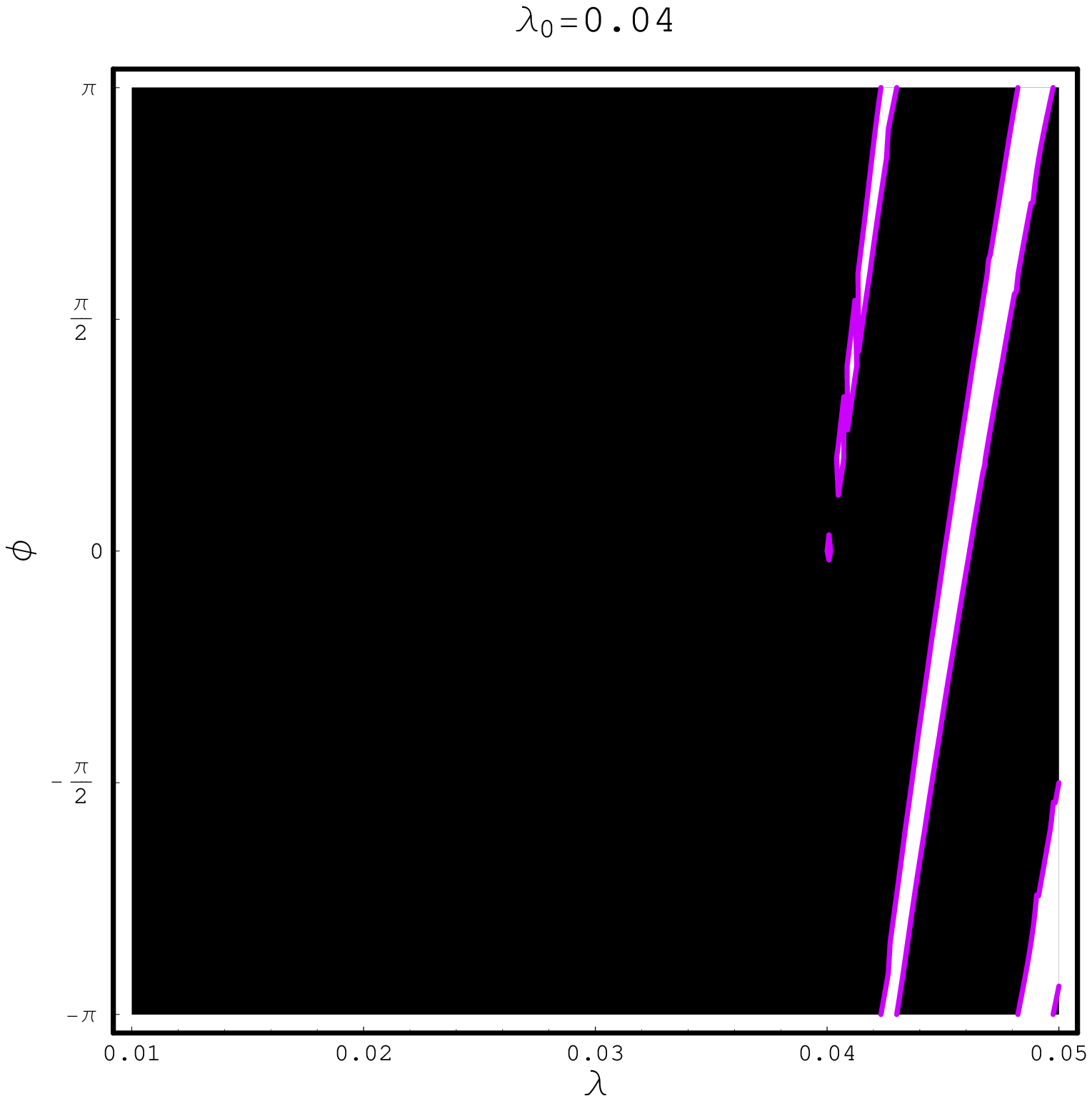} & \epsfxsize=2.8in
\epsffile{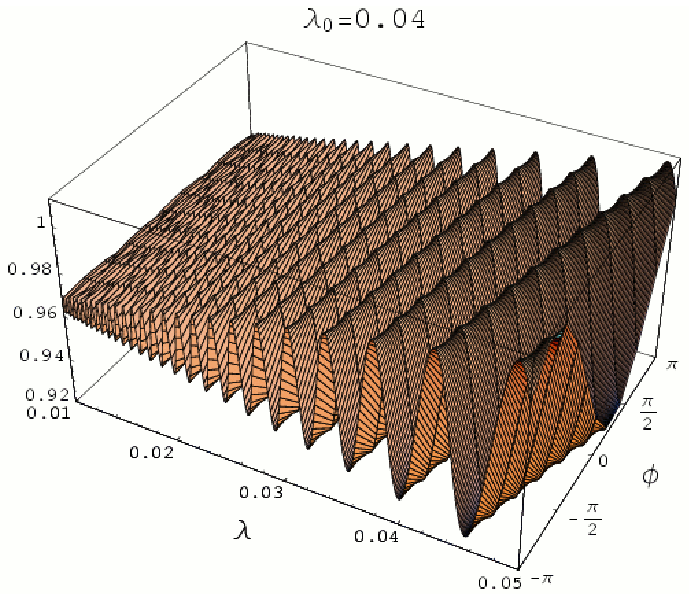}
\end{array}$
\end{center}
\caption{\label{fig:deger} (Left) The value of $I$ plotted on the
  $(\lambda, \phi)$ plane, where the purple (gray) lines, white areas and
  black areas represent $I=1$, $I>1$ and $I<1$ regions
  respectively. (Right) The 3--d representation of $I$. The underlying
  ``data'' has $\phi_0=0$ and $\epsilon_0=0.15/16=\epsilon$, and the
  underlying value of $\lambda_0$ ranges from (top to bottom) 0.005,
  0.01, 0.04 respectively.}
\end{figure*}

\begin{figure}
\begin{center}
$\begin{array}{c@{\hspace{1in}}c} \epsfxsize=2.8in
\epsffile{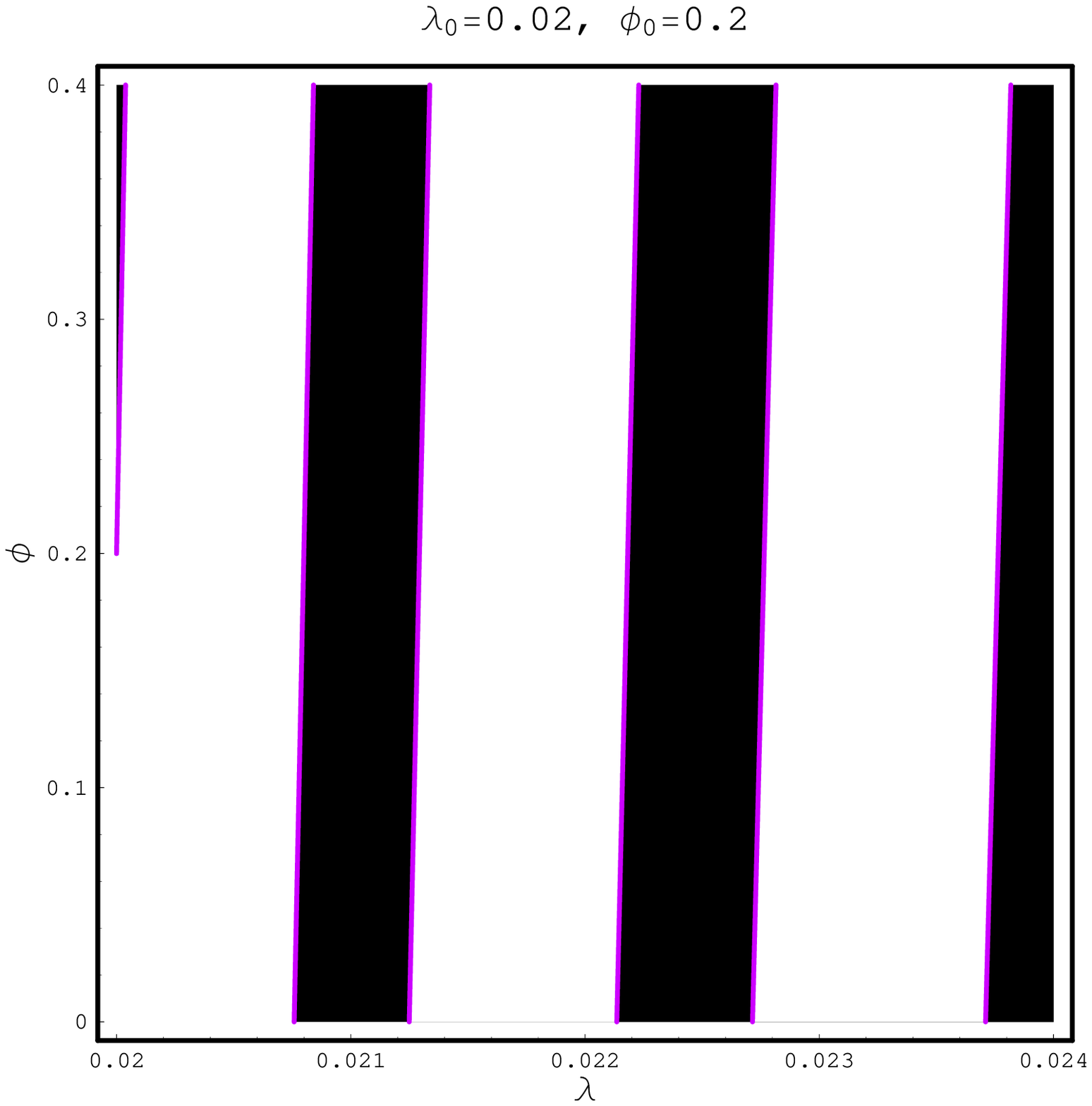} \\ \epsfxsize=2.8in
\epsffile{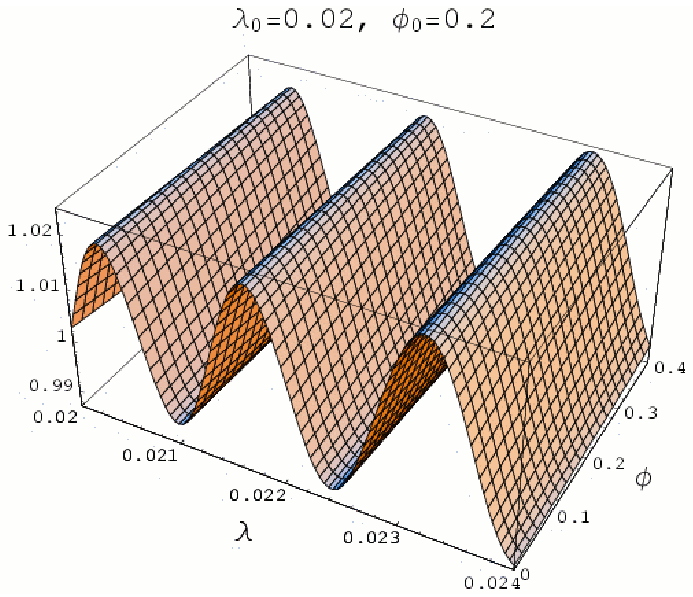}
\end{array}$
\end{center}
\caption{\label{fig:finegridI} $I$ plotted on the $(\lambda, \phi)$
  plane, showing in detail a small region for comparison with Figure
  \ref{fig:finegrid}. The underlying ``data'' has $\phi_0=0.2$,
  $\epsilon_0=0.15/16=\epsilon$, and $\lambda_0=0.02$.}
\end{figure}

Why are these degenerate ``islands'' present in the parameter space?
In  order to understand this result in a semi-analytic sense, we
consider the effect of the degeneracy on the primordial power spectra.
The motivation of the following analysis is that likelihood
degeneracies arise from frequency ``beating'' between underlying
modulations in the data and the model. Parameter combinations where
``beating'' occurs over the largest $k$--ranges give rise to discrete
likelihood ``islands'' or local maxima in the $(H/M)$ vs $\phi$ plane.

Defining the trans-Planckian modification factor multiplying the
standard Bunch-Davies power spectrum as
\begin{equation}
\label{eq:delta}
\Delta(k, \lambda, \phi, \epsilon) =
1+\lambda\left(\frac{k}{k_\star}\right)^{-\epsilon} \sin
\left[\frac{2}{1-\epsilon} \frac{1}{\lambda}
\left(\frac{k}{k_\star}\right)^\epsilon + \phi \right],
\end{equation}
where $\lambda \equiv (H/M)$, we define the parameter $I$ where
\begin{equation}
I = \frac{\int_{0.001}^{0.1} \Delta(k, \lambda, \phi,
  \epsilon)\Delta(k_0, \lambda_0, \phi_0, \epsilon_0)
  dk}{\int_{0.001}^{0.1} \Delta(k_0, \lambda_0, \phi_0, \epsilon_0)
  \Delta(k_0, \lambda_0, \phi_0, \epsilon_0) dk}.
\end{equation}
Here, we have performed the integration over the relevant scales for
the CMB, $k\sim 0.001 {\rm Mpc}^{-1}$ -- $0.1 {\rm Mpc}^{-1}$. If
$I=1$, there will be a near-perfect degeneracy between the parameter
sets $(\lambda_0, \phi_0, \epsilon_0)$ and $(\lambda, \phi,
\epsilon)$. Assuming that $\epsilon$ can be constrained by detecting
tensor modes, a local likelihood maximum is likely to be present at
$(\lambda, \phi)$ even though the underlying ``data'' has $(\lambda_0,
\phi_0)$.

In the case corresponding to our fiducial models above, $\lambda_0 = 0$,
only a simple degeneracy $1/\lambda \propto \pi n -\phi$ exists. In
Figures \ref{fig:deger}, we compute $I$ for several values of non-zero
$\lambda = [0.005, 0.01, 0.04]$. While the results in these figures
are in the analytically tractable power-spectrum space, Figure
\ref{fig:finegridI} shows a comparison with Figure \ref{fig:finegrid}
above, which was calculated the full $C_\ell$ formalism. Since
convolution with the CMB transfer functions will tend to smear out
likelihood degeneracies present in primordial power-spectrum space, we
expect the likelihood ``islands'' to be less prominent in the full
$C_\ell$ calculation than in the semi-analytic calculation.

These results show that the presence of discrete likelihood maxima in
parameter space gives some extra information beyond merely being a
numerical headache and a barrier to parameter estimation. Their
presence in a parameter estimation analysis is an indication of an
underlying modulation in the data, and the appearance of these
degeneracies  on the $(H/M)$ vs $\phi$ plane in  this model is a
function of the value of $(H/M)$ of the data. Of course, should such
an effect be detected in real data, whether or not such a modulation
is due to a primordial effect or some systematic effect in the data,
is a question which would need to be investigated thoroughly.

While the grid method is a good tool for exploring the global shape of
the likelihood function, it is an oversimplification to vary only two
parameters in the fit. Error bars derived from such a two-parameter
analysis will be unrealistically optimistic, since a complete analysis
must take into account degeneracies between the parameters of interest
and a large number of other cosmological parameters. To investigate
the effect of including other cosmological parameters in the fit, we
apply a Fisher matrix technique in Sec. \ref{sec:fishermatrix} and
Monte Carlo Markov chain methods in Sec. \ref{sec:markovchains}.

\subsection{Fisher Matrix}
\label{sec:fishermatrix}

The Fisher matrix technique is a simple and efficient method for
estimating the expected measurement errors when marginalizing over a
large number of parameters, for which grid methods would be
unacceptably time consuming.  Measurement uncertainty in cosmological
parameters is characterized by the Fisher information matrix
$\alpha_{ij}$. (For a review, see Ref. \cite{Tegmark:1996bz}.) Given a
set of parameters $\left\lbrace \lambda_i \right\rbrace$, the Fisher
matrix is given by
\begin{equation}
\alpha_{ij} = \sum_l \sum_{X,Y} {\partial C_{Xl} \over \partial
\lambda_i} {\rm Cov}^{-1}\left(\hat C_{Xl} \hat C_{Yl}\right)
{\partial C_{Yl} \over \partial \lambda_j},
\end{equation}
where $X,Y = T,E,B,C$ and $\rm{Cov}^{-1}\left(\hat C_{Xl} \hat
C_{Yl}\right)$ is the inverse of the covariance matrix between the
estimators $\hat C_{Xl}$ of the power spectra. Calculation of the
Fisher matrix requires assuming a ``true'' set of parameters and
numerically evaluating the $C_{Xl}$'s and their derivatives relative
to that parameter choice. The covariance matrix for the parameters
$\left\lbrace \lambda_i\right\rbrace$ is just the inverse of the
Fisher matrix, $\left(\alpha^{-1}\right)_{ij}$, and the expected error
in the parameter $\lambda_i$ is of order
$\sqrt{\left(\alpha^{-1}\right)_{ii}}$.  The full set of parameters
$\left\lbrace \lambda_i \right\rbrace$ we allow to vary is:
\begin{enumerate}  
\item{tensor/scalar ratio $r$,}
\item{spectral index $n_s$,}
\item{running $dn_s/d\ln k$,}
\item{normalization $A(k_* = 0.002\ {\rm Mpc}^{-1})$,}
\item{baryon density $\Omega_b,$}
\item{Hubble constant $h \equiv H_0 / (100\,{\rm
km\,sec^{-1}\,Mpc^{-1}})$,}
\item{reionization optical depth, $\tau_{\rm ri}$,}
\item{Cosmological constant density $\Omega_\Lambda$,}
\item{Trans-Planckian amplitude $H/M$,}
\item{Trans-Planckian phase $\phi$.}
\end{enumerate}
The total density is fixed such that the universe is flat,
$\Omega_b + \Omega_c + \Omega_\Lambda = 1$. Since
one needs only to calculate the derivative of the likelihood function
with respect to each parameter, Fisher matrix calculations scale
linearly with the number of parameters and are therefore very
efficient for forecasting errors in fits marginalizing over a large
number of parameters. This  technique, however, gives no information
about the {\em global} shape of the likelihood surface. Obviously,
since we do not know the values of the trans-Planckian parameters in
the real universe, only the size of the errors, not the location of
the central value, is significant.

We perform a Fisher matrix analysis for two different forecast data
sets: first, in order to understand the best-case scenario, we
consider a measurement for which the errors are limited by cosmic
variance to $\ell = 1500$, similar to that assumed for the grid method
in Sec. \ref{sec:grid}.  Second, we consider the more immediately
realistic case of a measurement with error bars similar to those
projected for the Planck Surveyor satellite, for which we assume a
two-channel experiment measuring both temperature and
polarization. The 143 GHz channel is assumed to have an $8$-arcminute
beam and a  pixel sensitivity of $\sigma_{\rm pix} = 5.5\ {\rm \mu
K}$, and the 217 GHz channel is assumed to have an $5.5$-arcminute
beam and a  pixel sensitivity of $\sigma_{\rm pix} = 11.7\ {\rm \mu
K}$.
 
\begin{figure*}
\begin{center}
$\begin{array}{c@{\hspace{1in}}c} \epsfxsize=2.4in
\epsffile{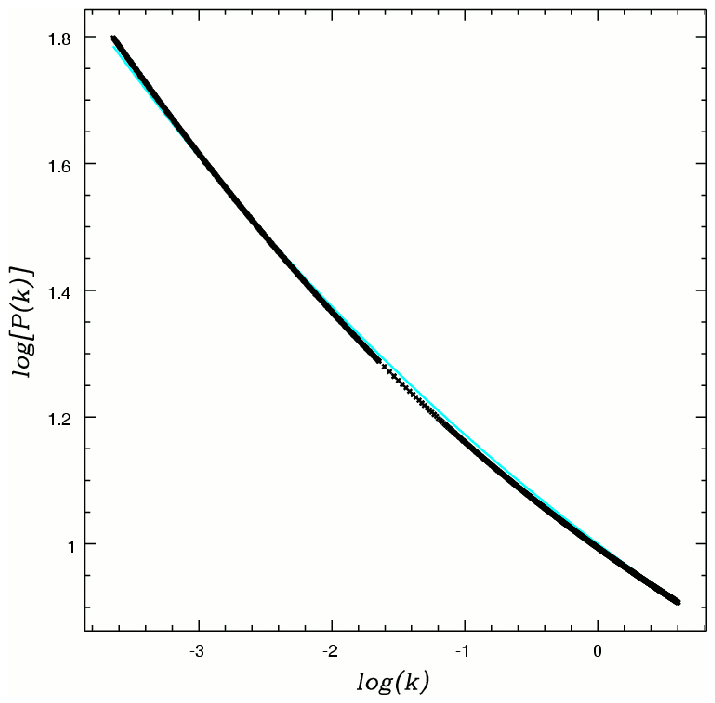} & \epsfxsize=2.4in
\epsffile{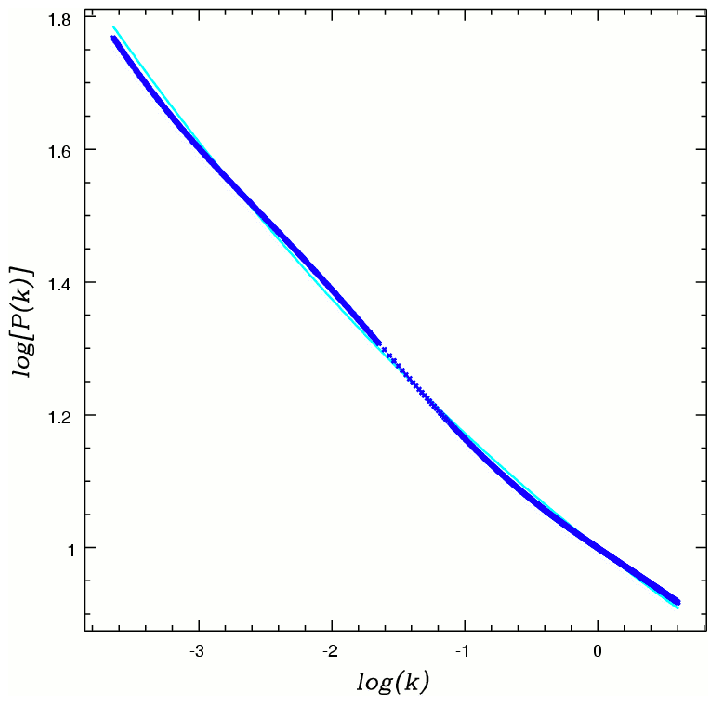} \\  \epsfxsize=2.4in
\epsffile{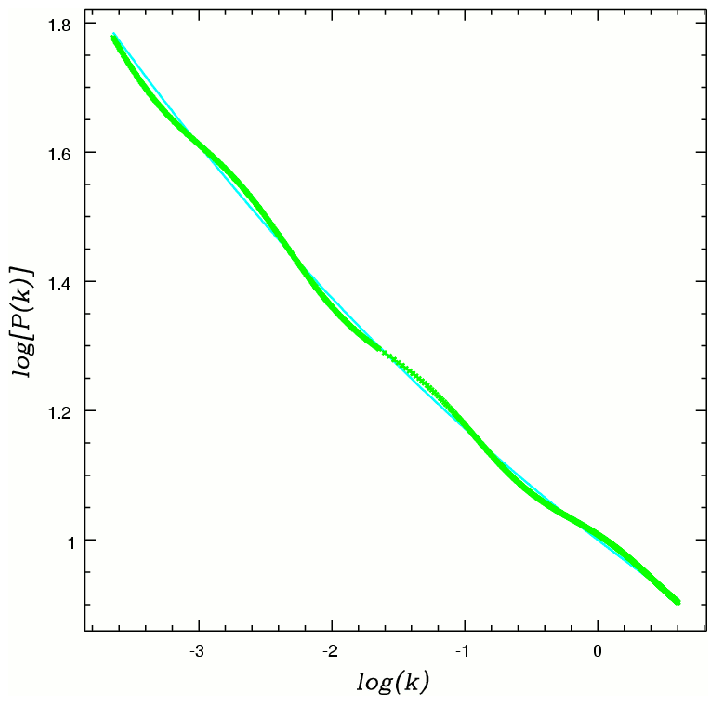} & \epsfxsize=2.4in
\epsffile{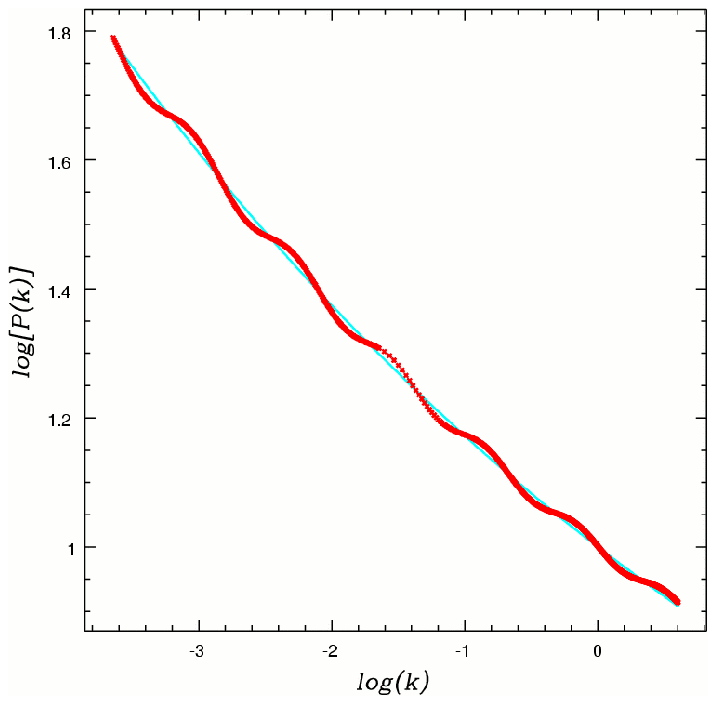} \\
\end{array}$
\end{center}
\caption{\label{fig:modulation}Modulation of the power spectrum for
various choices of $r$: $r = 0.02$ (top left), $r = 0.05$ (top right),
$r = 0.1$, (bottom left), and  $r = 0.2$ (bottom right). In all cases
the solid (cyan) line is the unmodulated  spectrum, and the points
represent the modulated power spectrum as sampled by  CMBFAST.}
\end{figure*}

We are especially interested in the degeneracy between trans-Planckian
modulations and the other parameters determining the shape of the
primordial power spectrum, $n_s$ and $dn_s/d\ln k$. This degeneracy is
highly dependent on the value of $r$ in the underlying fiducial model
we choose, because for small $r$, the trans-Planckian modulation
becomes very long-wavelength and thus mimics a change in the spectral
index $n_s$ or running $dn_s/d\ln k$ (Fig. \ref{fig:modulation}). (For
the case $r = 0$, the ``modulation''  loses all scale-dependence and
simply affects the normalization. The value of $H/M$ thus decouples
from the  other parameters determining the shape of the power
spectrum, and the degeneracy  vanishes in this limit.)

\begin{figure}
\includegraphics[width=3.1in]{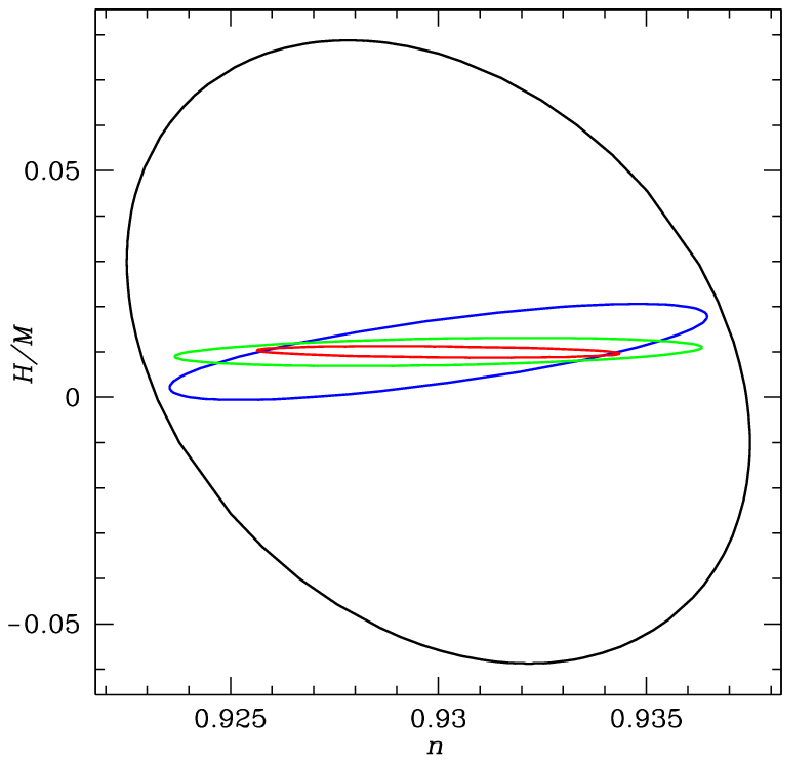}
\includegraphics[width=3.1in]{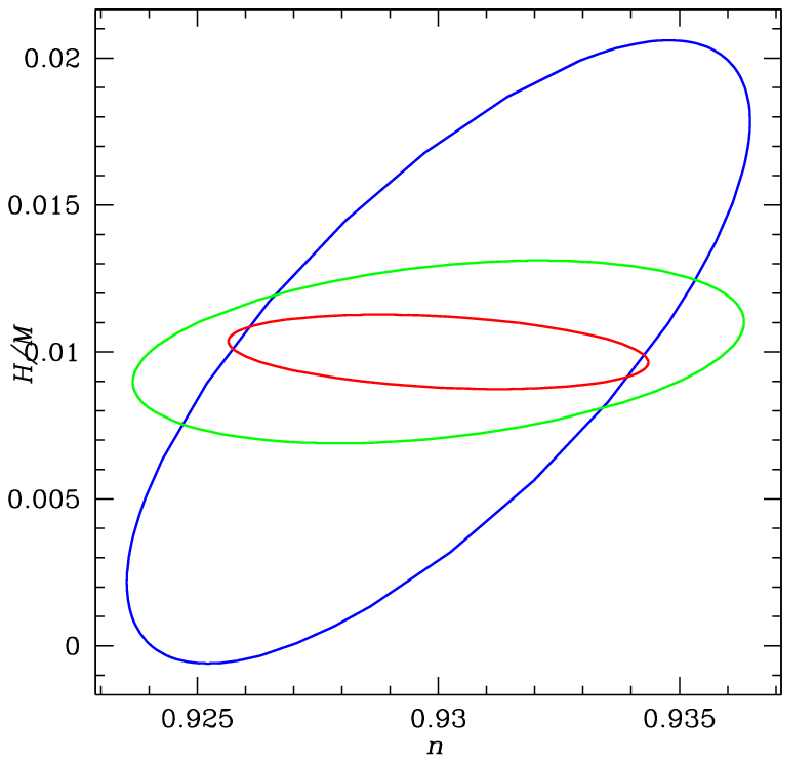}
\caption{\label{fig:cosvarfisher} $2\sigma$ error ellipses in the $n_s$
  - $H/M$ plane, calculated by Fisher matrix, assuming a fiducial
  model with  $n_s = 0.93$ and $H/M$ = 0.01 and a cosmic-variance
  limited measurement to $\ell = 1500$. The different ellipses show
  the effect of changing the choice of tensor/scalar ratio $r$ in the
  fiducial model: larger values of $r$ make it easier to distinguish
  the trans-Planckian modulation. The top panel shows ellipses for $r
  = 0.02$ (black, outer), $r = 0.05$ (blue, next in), $r = 0.1$
  (green, second smallest), and $r = 0.2$ (red, inner). The bottom
  panel shows the same figure zoomed in to show the three inner
  ellipses. (Negative values for $H/M$ are nonphysical, and are
  included to show the full error ellipse.)}
\end{figure}

\begin{figure}
\includegraphics[width=3.1in]{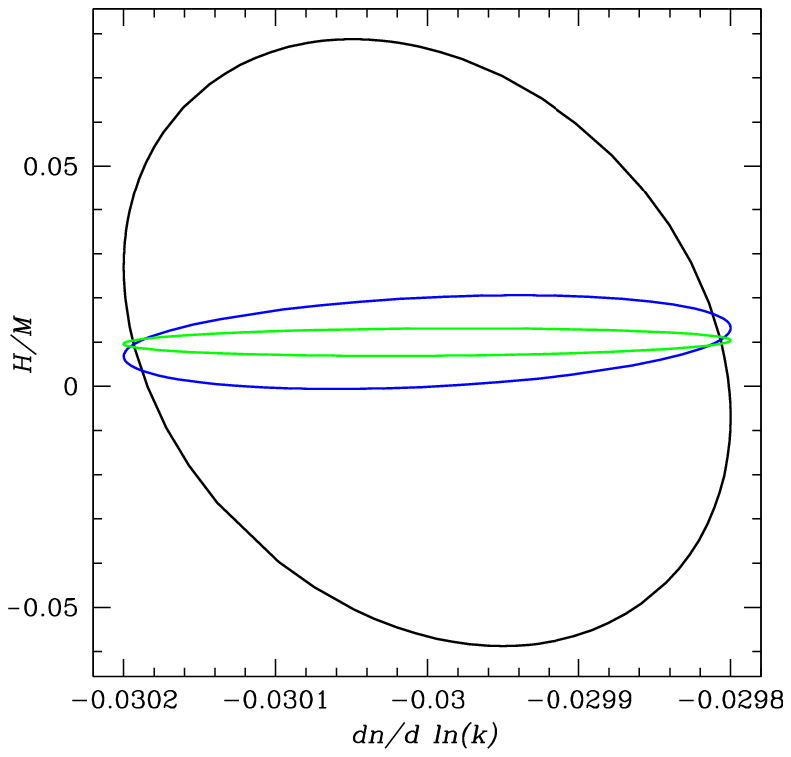}
\includegraphics[width=3.1in]{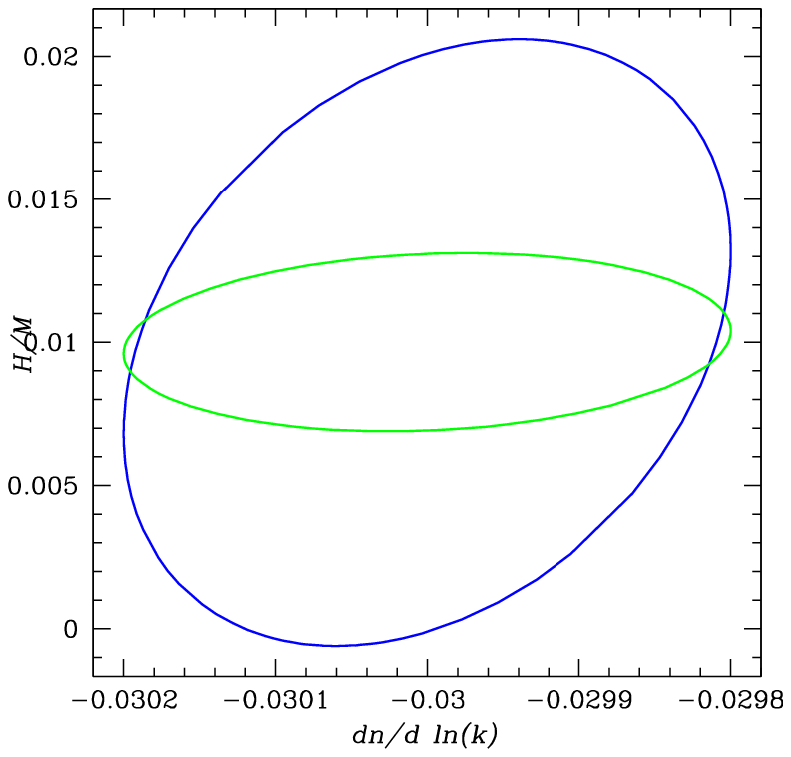}
\caption{\label{fig:cosvarfisherdn} $2\sigma$ error ellipses in the
$dn_s/d\ln k$ - $H/M$ plane, calculated by Fisher matrix, assuming a
fiducial model with  $n_s = 0.93$ and $H/M$ = 0.01 and a 
cosmic-variance limited observation. The
different ellipses show the effect of changing the choice of
tensor/scalar ratio $r$ in the fiducial model: larger values of $r$
make it easier to distinguish the trans-Planckian modulation. The top
panel shows ellipses for $r = 0.02$ (black, outer), $r = 0.05$ (blue,
next in), and $r = 0.1$ (green, smallest). The bottom panel shows 
the same figure zoomed in to show the three inner ellipses.}
\end{figure}

We therefore plot error ellipses in the $n_s$ - $H/M$ plane and the
$dn_s/d\ln k $ - $H/M$ plane for various choices for $r$ running from $r
= 0.02$ to $0.2$. The fiducial model assumed in all cases has $H/M =
0.01$ and phase $\phi = 0$. Other parameters are identical to those
used for the grid calculation in Sec. \ref{sec:grid}. Figure
\ref{fig:cosvarfisher} shows the $n_s$ - $H/M$ plane for the case of the
cosmic-variance limited measurement, and Fig. \ref{fig:cosvarfisherdn} 
shows the $dn_s/d\ln k$ - $H/M$ plane. In this case, the measurement is
sufficiently accurate that the degeneracy between the power spectrum
parameters $n_s$, $dn_s/d\ln k$ and the trans-Planckian amplitude $H/M$ is  
negligible, although the ability to constrain $H/M$ is significantly degraded 
for small values of $r$. These error estimates can be directly compared
with those generated by a full Markov Chain Monte Carlo analysis with
similar parameters,  (Figs. \ref{fig:params_bigr} and
\ref{fig:params_tinyr}), and are in good agreement.
\begin{figure}
\includegraphics[width=3.1in]{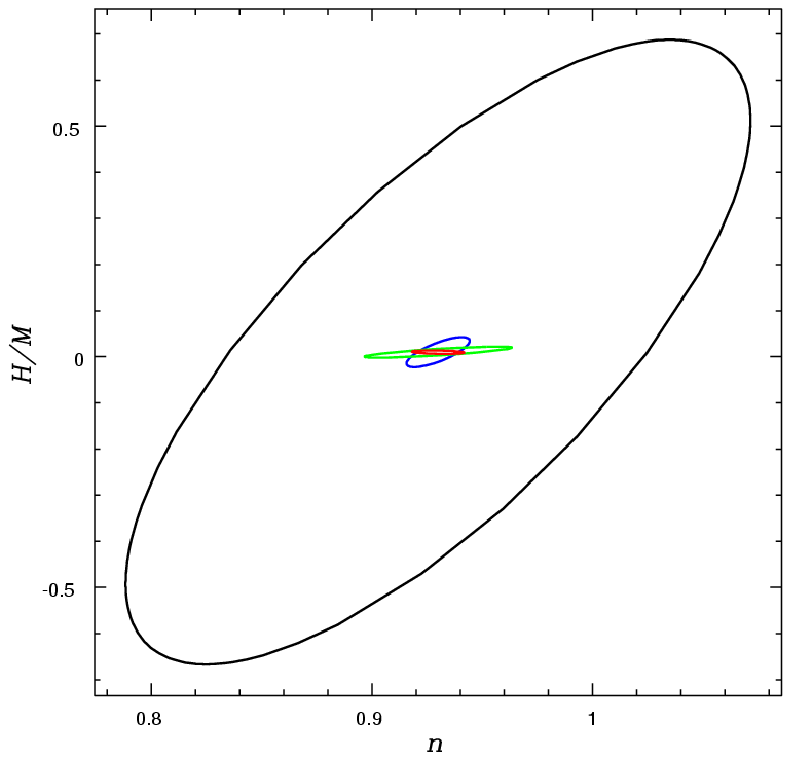}
\includegraphics[width=3.1in]{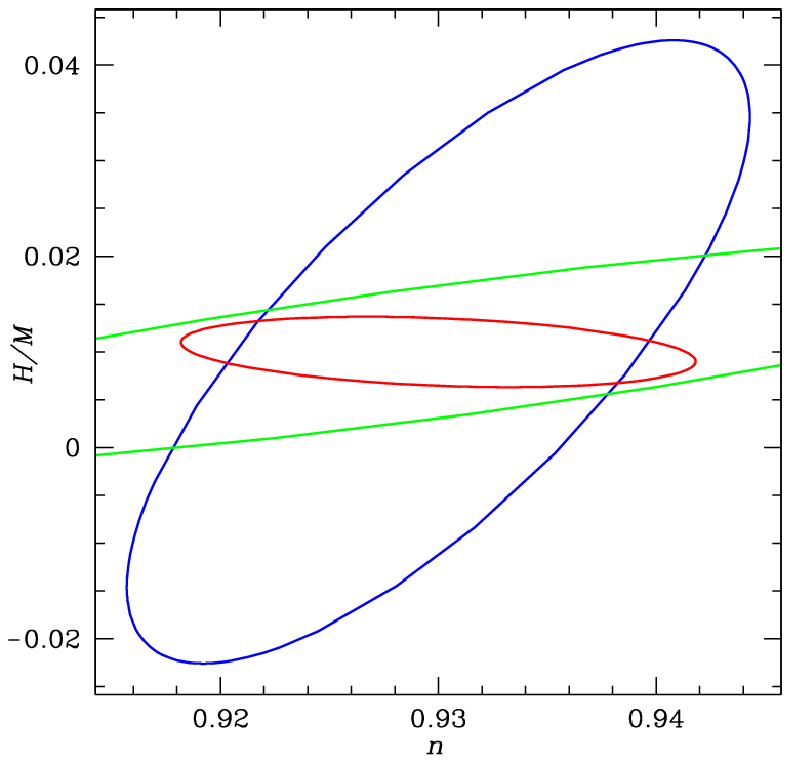}
\caption{\label{fig:Planckfishern} $2\sigma$ error ellipses in the $n_s$
  - $H/M$ plane, calculated by Fisher matrix, assuming a fiducial
  model with $n_s = 0.93$ and $H/M$ = 0.01 and an observation with the
  projected sensitivity of the Planck Surveyor satellite. The
  different ellipses show the effect of changing the choice of
  tensor/scalar ratio $r$ in the fiducial model: larger values of $r$
  make it easier to distinguish the trans-Planckian modulation. The
  top panel shows ellipses for $r = 0.02$ (black, outer), $r = 0.05$
  (blue, next in), $r = 0.1$ (green, second smallest), and $r = 0.2$
  (red, inner). The bottom panel shows the same figure zoomed in to
  show the three inner ellipses.}
\end{figure}
\begin{figure}
\includegraphics[width=3.1in]{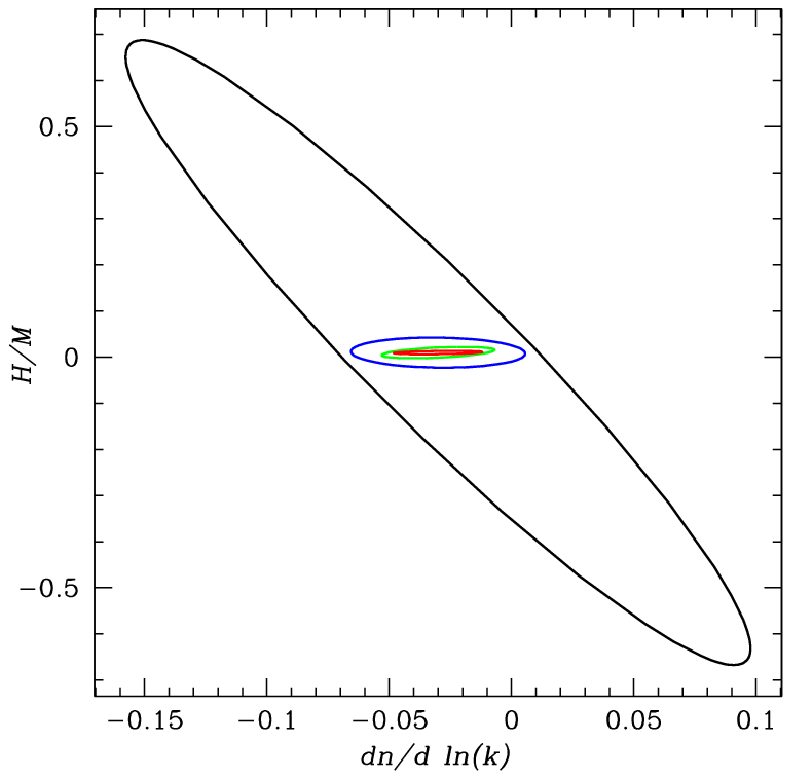}
\includegraphics[width=3.1in]{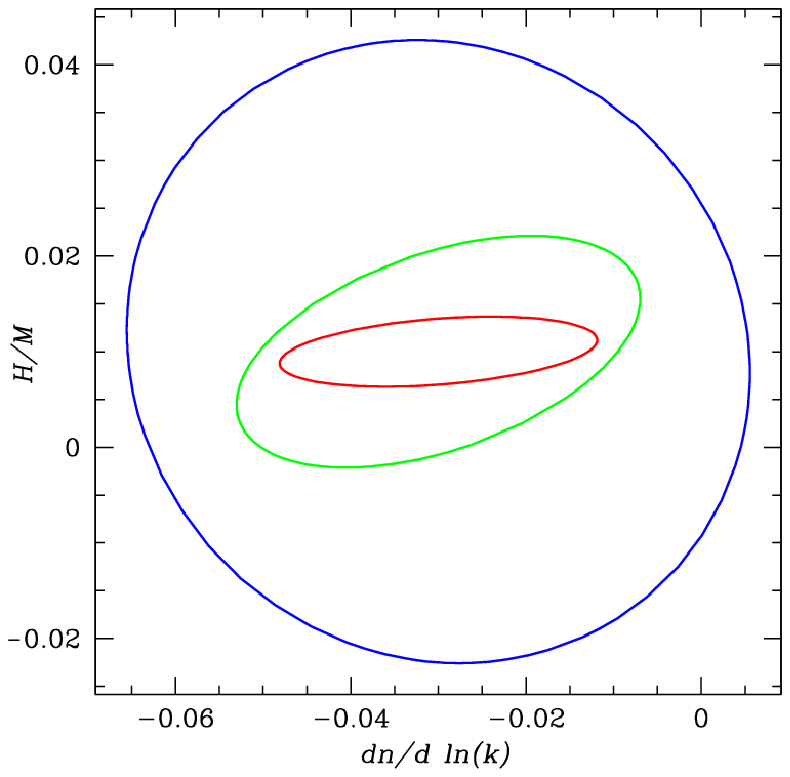}
\caption{\label{fig:Planckfisherdn} $2\sigma$ error ellipses in the
$dn_s/d\ln k$ - $H/M$ plane, calculated by Fisher matrix, assuming a
fiducial model with  $n_s = 0.93$ and $H/M$ = 0.01 and an observation
with the projected sensitivity of the Planck Surveyor satellite. The
different ellipses show the effect of changing the choice of
tensor/scalar ratio $r$ in the fiducial model: larger values of $r$
make it easier to distinguish the trans-Planckian modulation. The top
panel shows ellipses for $r = 0.02$ (black, outer), $r = 0.05$ (blue,
next in), $r = 0.1$ (green, second smallest), and $r = 0.2$ (red,
inner). The bottom panel shows the same figure zoomed in to show the
three inner ellipses.}
\end{figure}
This situation is very different for the case of Planck Surveyor,
shown in  Figs.  \ref{fig:Planckfishern} and \ref{fig:Planckfisherdn}
for the $n_s$ - $H/M$ and $dn_s/d\ln k$ - $H/M$ planes, respectively.  We
see that in this case, the inclusion of a trans-Planckian modulation
in the parameter set strongly degrades the ability to measure the
power spectrum parameters. This degeneracy is lifted if the
tensor/scalar ratio is large enough, effectively disappearing
altogether for a tensor/scalar ratio of order $r \sim 0.1$.

Finally, as expected, the measurement uncertainties on the
trans-Planckian parameters increase significantly when a full set of
other cosmological parameters is taken  into account. The Fisher
matrix analysis is in good agreement with the full Markov Chain Monte
Carlo analysis discussed in Sec. \ref{sec:markovchains}, indicating
that a trans-Planckian modulation of order $H/M = 0.01$ is only
detectable at the $2\sigma$ level in the most optimistic case of a
very large tensor/scalar ratio $r \sim 0.1$. In the next section we
discuss the MCMC analysis in detail and derive the error  ellipses
relative to a fiducial model with $H/M = 0$, testing our ability to
distinguish a signal from the null hypothesis. This analysis (which
includes a longer lever arm of $\ell = 2000$), indicates that a
cosmic-variance limited measurement can place a $2\sigma$ upper limit
on  trans-Planckian fluctuations of around $H/M = 0.004$ for the case
with a large tensor/scalar ratio ($r = 0.15$), and $H/M = 0.03$ for the 
case with an undetectably small tensor/scalar ratio.

\section{Markov Chains} 
\label{sec:markovchains}

\subsection{Introduction}

In this section, we use a Markov Chain Monte Carlo (MCMC) technique to
evaluate the likelihood function of model parameters. This approach,
proposed by \cite{Christensen:2000ji} in the context of cosmological
parameter estimation, has become the standard tool for such analyses
\citep[e.g.][]{Christensen:2001gj,Knox:2001fz,Lewis:2002ah,Kosowsky:2002zt,Verde:2003ey}.
MCMC is a method to simulate posterior distributions. In particular,
we simulate observations from the posterior distribution ${\cal
P}({\bf \alpha}|x)$, of a set of parameters ${\bf \alpha}$ given event
$x$, obtained via Bayes' Theorem,
\begin{equation}
{\cal P}(\alpha|x)=\frac{{\cal P}(x|\alpha){\cal P}(\alpha)}{\int
{\cal P}(x|\alpha){\cal P}(\alpha)d\alpha},
\label{eq:bayes}
\end{equation}
\noindent where ${\cal P}(x|\alpha)$ is the likelihood of event $x$
given the model parameters $\alpha$ and ${\cal P}(\alpha)$ is the
prior probability density. The MCMC generates random draws
(i.e. simulations) from the posterior distribution that are a ``fair''
sample of the likelihood surface. From this sample, we can estimate
all of the quantities of interest about the posterior distribution
(mean, variance, confidence levels). The MCMC method scales
approximately linearly with the number of parameters, thus allowing us
to perform a likelihood analysis in a reasonable amount of time.

A properly derived and implemented MCMC draws from the joint posterior
density ${\cal P}(\alpha|x)$ once it has converged to the stationary
distribution. For our application, $\alpha$ denotes a set of 10
parameters. The set is almost the same as used for the Fisher matrix
calculation. We have 4 ``late-time'' parameters: the physical energy
density in baryons $\Omega_b h^2$, the total physical energy density
in matter $\Omega_m h^2$, the optical depth to reionization $\tau_{\rm ri}$,
the Hubble constant in units of $100$ km s$^{-1}$ Mpc -- $h$.  There
are   6 ``primordial parameters'': the tensor-to-scalar ratio $r$, the
spectral slope $n_s$, the running of the spectral slope $d n_s/d \ln
k$, the power spectrum amplitude $A$, and the parameters of the
trans-Planckian model defined above, $H/M$ and $\phi$. Event $x$ will
be the set of observed $\widehat{\cal C}_{\ell}$. The priors on the
model are that the universe is flat, so that $\Omega_m +
\Omega_\Lambda = 1$, and the dark energy is a cosmological constant
with equation of state $p=-\rho$. The definitions used for the
unmodified primordial power spectra of the scalar and tensor
perturbations $\Delta^2_{\cal R, BD}(k) \equiv
k^3/(2\pi^2)\langle\left|{\cal R}_{\mathbf k}\right|^2\rangle\propto
k^{n_s-1}$ and $\Delta^2_{h, BD}(k) \equiv 2
k^3/(2\pi^2)\langle\left|h_{+{\mathbf k}}\right|^2 +
\left|h_{\times{\mathbf k}}\right|^2\rangle\propto k^{n_t}$ are given
in Section \ref{sec:model}.

\subsection{Methodology}

The experiment in the case study is considered to measure ${TT, TE,
EE, BB}$ power spectra to $\ell=2000$; beyond this, the systematic
errors in removing the contamination from secondary CMB anisotropies
are likely to be non-negligible. We include the effect of weak lensing
on all power spectra, since this is the ``theoretical limit'' in the
accuracy of the measurement - weak lensing aliases power between
multipoles and tends to smooth out small-scale oscillations in the
data. Therefore, any of the further uncertainties that must be
accounted for in a practical CMB experiment, such as foreground
removal, partial sky coverage etc, will degrade the constraints
obtainable on the trans-Planckian parameters relative to those found
in this study.

For the simulated ``data'', we consider two cases: one with
significant primordial tensor modes (denoted Case H), and one with an
undetectably small contribution of primordial tensors (denoted case
L). Case H is a mass-term inflation model with tensor-to-scalar ratio
$r=0.15$ which is consistent with {\sl WMAP} data at the $1\sigma$
level. Case L is simply the {\sl WMAP} LCDM concordance model with
$r=0.00013$ (quadrupole ratio $C_2^T/C_2^S \sim 6\times10^{-5}$).

The objective of this study is to see how far from  the null case the
trans-Planckian parameters must be before they are distinguished from
the null model by at least $2\sigma$. Consequently, these fiducial
models do not contain any trans-Planckian oscillations. Our fiducial
models also do not contain any running of the primordial scalar
spectral index; however, the effect of this parameter would be
degenerate with a very long wavelength trans-Planckian
modulation. Hence we marginalize over this parameter in our analysis.

In the practical implementation of the MCMC, we ran 16 chains for Case
H and 32 chains for Case L, each started at random $3\sigma$ steps
away from the fiducial model. By making use of preliminary test
chains, the proposal distribution was calibrated using the covariance
between the parameters in order to minimize correlations; the chains
were configured to step in the principal directions of the parameter
covariance matrix optimal $\sim 24$\% acceptance rates for case H; for
case L, only a few \% acceptance rates were achieved even after
several recalibrations of the proposal matrix, indicating the
complexity of the likelihood surface. The chains were run on a
high-performance 64-bit Opteron Beowulf cluster, and 300,000 and
500,000 total models were computed for Case H and Case L,
respectively.   The presence of ``islands'' in the likelihood space
meant that the chains converged relatively slowly, compared to what is
typical for a fit to CMB observations when the trans-Planckian terms
are absent.  In fact, we stopped the chains for Case L slightly short
of full convergence, but long after the predicted parameter regions
had stabilized.

\subsection{The Shape of the Likelihood Surface}

\begin{figure*}
\includegraphics[width=7.1in]{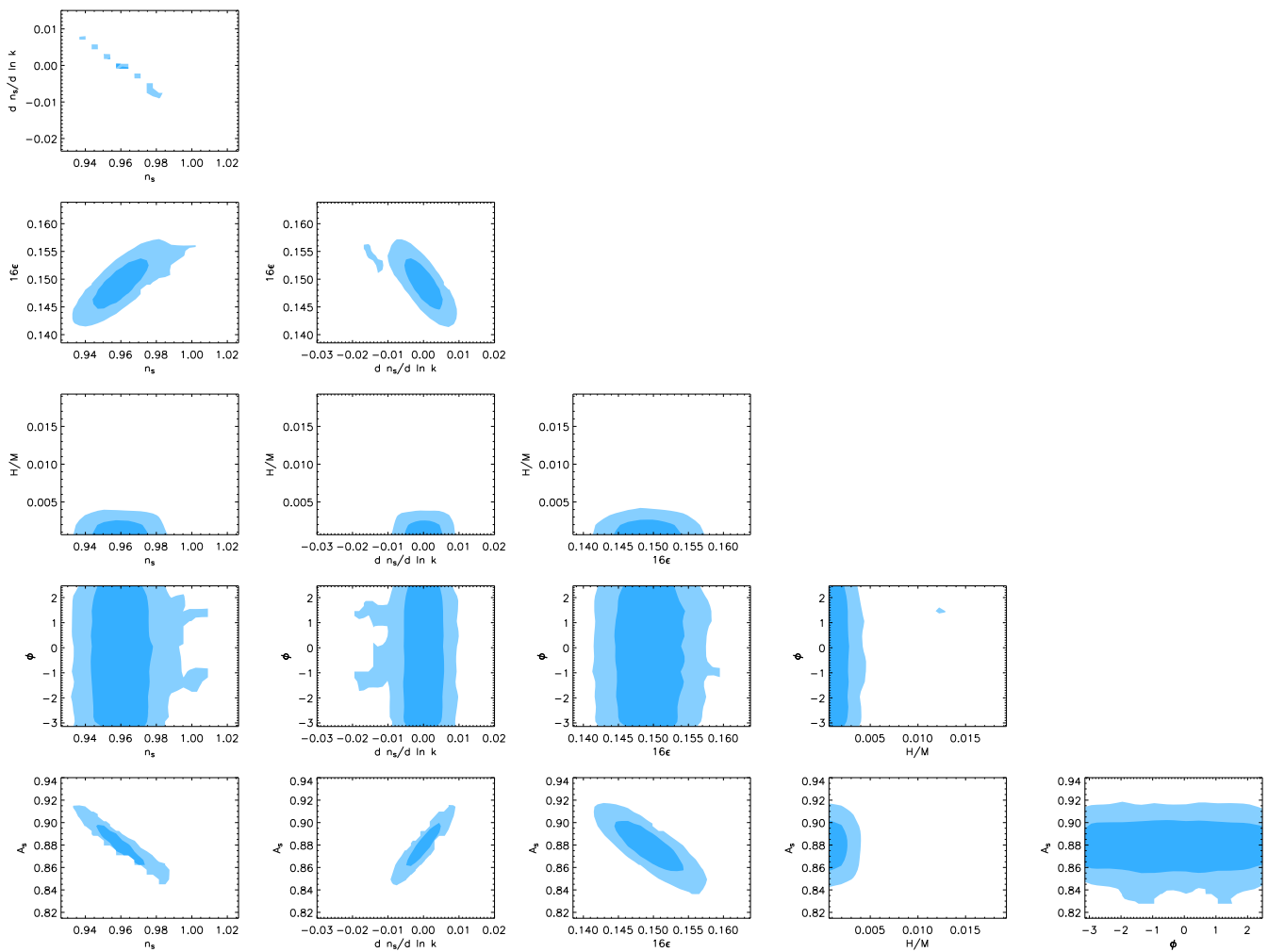}
\caption{\label{fig:params_bigr} Case H: A fiducial model with
  $r=0.15$ and no trans-Planckian modulations. The 2-d joint 68\%
  (dark) and 95\% (light) confidence regions for pairs of parameters
  in the set \{$r$, $n_s$, $d n_s/d \ln k$, $A$, $H/M$, $\phi$\}
  obtained via MCMC methods. In the case of each parameter pair, all
  the other eight parameters in the model have been marginalized
  over. The discrete ``blobs'' in the ($n_s$, $d n_s/d \ln k$) plane
  are not physical but a graphical effect arising as consequence of
  binning of a narrow likelihood degeneracy.
  }
\end{figure*}

\begin{figure*}
\includegraphics[width=7.1in]{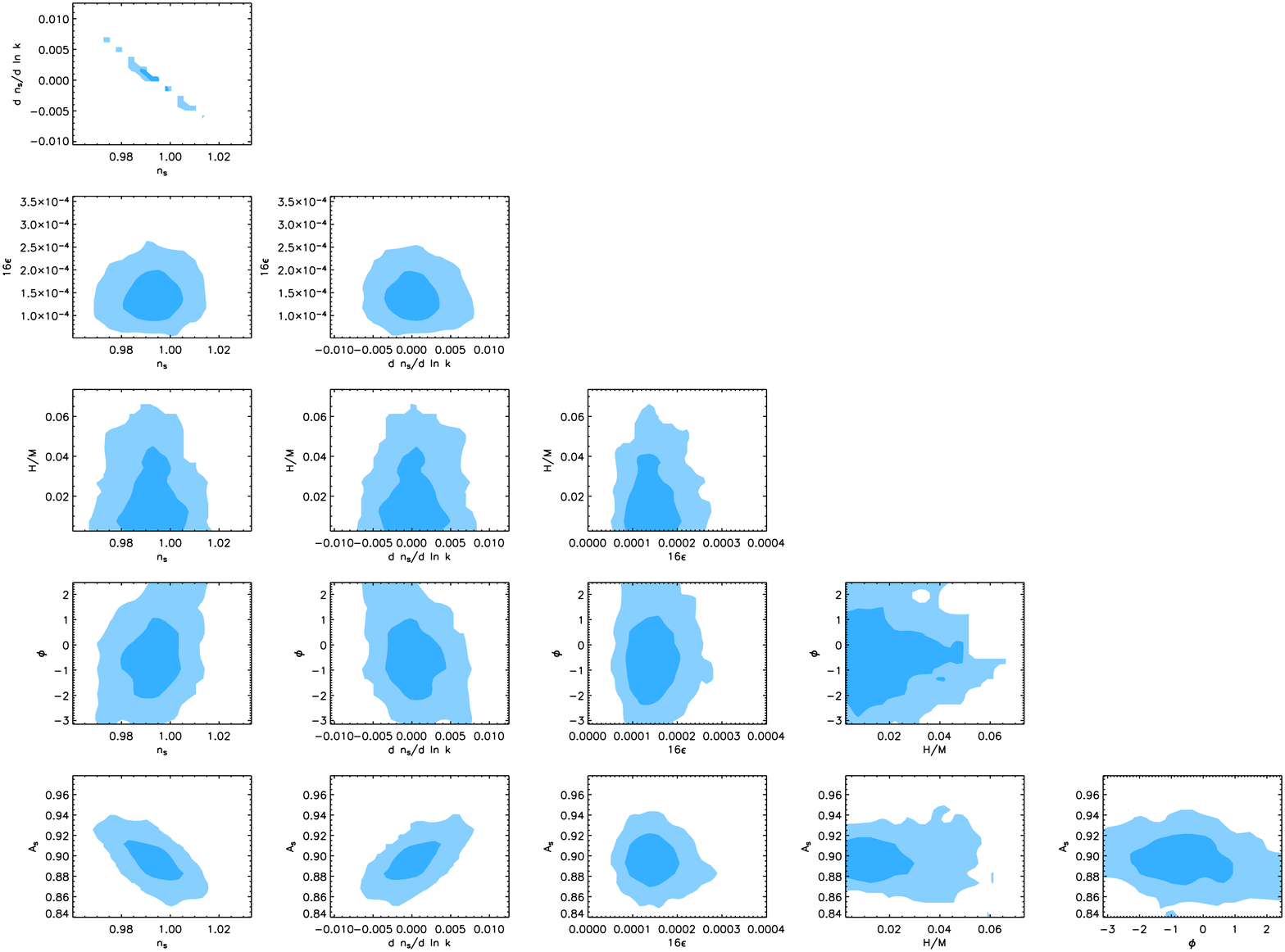}
\caption{\label{fig:params_tinyr} Case L: A fiducial model with
  $r=0.00013$ and no trans-Planckian modulations. The 2-d joint 68\%
  (dark) and 95\% (light) confidence regions for pairs of parameters
  in the set \{$r$, $n_s$, $d n_s/d \ln k$, $A$, $H/M$, $\phi$\}
  obtained via MCMC methods. In the case of each parameter pair, all
  the other eight parameters in the model have been marginalized
  over. The discrete ``blobs'' in the ($n_s$, $d n_s/d \ln k$) plane
  are not physical but a graphical effect arising as consequence of
  binning of a narrow likelihood degeneracy.}
\end{figure*}

Tables \ref{table:params_bigr} and \ref{table:params_tinyr} show the
fiducial models and recovered marginalized 1--d 68\% probability
parameter constraints on the ten model parameters for Case H and Case
L, respectively. Figures \ref{fig:params_bigr} and
\ref{fig:params_tinyr} show the marginalized 2--d joint 68\% and 95\%
confidence regions for pairs of the six ``primordial'' parameters for
the same models. In Case H, which has potentially observable
primordial tensor modes, a deviation of $H/M$ from null is detectable
at greater than $2\sigma$ level if $(H/M)>0.004$ in an ideal
experiment, which renders $H/M \sim 0.01$ potentially detectable in a
real-world experiment. In Case L, which has unobservably small
primordial tensor modes, one requires $(H/M)>0.03$ in order to be
detectable at greater than the $2\sigma$ level in an ideal experiment,
rendering such a detection implausible in the real
world. Ref. \cite{Okamoto:2003wk} discussed a degeneracy between
$\epsilon$ and $(H/M)$ which appears to be essentially eliminated in
our model due to the extra constraint on $\epsilon$ from tensor
modes.

\begin{table}
\caption{\footnotesize \label{table:params_bigr}  Case H: A fiducial
 model with $r=0.15$ and no trans-Planckian modulations. The input
 model, recovered mean parameters and the 68\% probability level of
 the 1--d marginalized likelihood. The derived parameter $\sigma_8$
 (the linear theory prediction for the amplitude of fluctuations
 within 8 Mpc/h spheres) is given in the last row.}
\begin{tabular}{lcc}
\hline \hline Parameter & Input & Recovered\\ \hline $\Omega_b h^2$ &
$0.02273$ & $0.02273 \pm 0.00006$\\ $\Omega_m$ & $0.3097$ & $0.3096
\pm 0.0016$\\ $h$  & $0.684$ & $0.684 \pm 0.001$ \\ $\tau_{\rm ri}$ & $0.112$ &
$0.112 \pm 0.002$ \\ $n_s(k_\star=0.002\ {\rm Mpc}^{-1})$ & $0.962$ &
$0.964 \pm 0.013$ \\ $d n_s/ d\ln k$ & $0$ & $0.000 \pm 0.004$ \\
$A(k_\star=0.002\ {\rm Mpc}^{-1})$ & $0.879$ & $0.878 \pm 0.016$\\
$r(k_\star=0.002\ {\rm Mpc}^{-1})$ & $0.150$ & $0.150 \pm 0.004$ \\
$H/M$ & $0$ & $0.002 \pm 0.002$\\ $\phi$ & $0$ & (degeneracy; mean
0.02) \\ $\sigma_8$ & & $0.871 \pm 0.001$ \\ \hline
\end{tabular}
\end{table}

\begin{table}
\caption{\footnotesize \label{table:params_tinyr}  Case L: A fiducial
 model with $r=0.00013$ and no trans-Planckian modulations. The input
 model, recovered mean parameters and the 68\% probability level of
 the 1--d marginalized likelihood. The derived parameter $\sigma_8$
 (the linear theory prediction for the amplitude of fluctuations
 within 8 Mpc/h spheres) is given in the last row.}
\begin{tabular}{lcc}
\hline \hline Parameter & Input & Recovered\\ \hline $\Omega_b h^2$ &
$0.02380$ & $0.02379 \pm 00006$\\ $\Omega_m$ & $0.2943$ & 0.$2945 \pm
0.0016$ \\ $h$  & $0.700$ & $0.700 \pm 0.001$\\ $\tau_{\rm ri}$ & $0.166$ &
$0.166 \pm 0.002$ \\ $n_s(k_\star=0.002\ {\rm Mpc}^{-1})$ & $0.994$ &
$0.993 \pm 0.010$ \\ $d n_s/ d\ln k$ & $0$ & $0.000 \pm 0.003$ \\
$A(k_\star=0.002\ {\rm Mpc}^{-1})$ & $0.900$ & $0.900 \pm 0.020$\\
$r(k_\star=0.002\ {\rm Mpc}^{-1})$ & $0.00013$ & $0.00014 \pm
0.00004$\\ $H/M$ & $0$ & $0.02 \pm 0.02$ \\ $\phi$ & $0$ &
(degeneracy; mean 0.05)\\ $\sigma_8$ & & $0.927 \pm 0.001$\\ \hline
\end{tabular}
\end{table}

The key observations to be drawn from these results are:
\begin{enumerate}
\item The   constraint on $H/M$ differs by an order of magnitude
  between Case H and Case L. This reinforces the point made above that
  {\em if\/} our Universe happens to be in a favorable part of the
  parameter space where the energy scale of inflation is high enough
  to produce an observable amplitude of primordial tensor modes, then
  one has a significantly improved chance of detecting any
  trans-Planckian modifications to the spectrum whose properties are
  similar to those of the case tested here.
\item The cosmological parameters are constrained at the $\sim 1\%$
  level except for the scalar spectral index and the amplitude of
  fluctuations at $k=0.002\ {\rm Mpc}^{-1}$, which are only
  constrained at the $\sim 10\%$ level. Therefore, another
  cosmological data-set that aids in constraining these parameters
  (for example, the proposed SKA survey, \cite{Rawlings:2004wk}) will
  minimize the degeneracies with these parameters shown in
  Figs.\ref{fig:params_bigr} and \ref{fig:params_tinyr} and improve
  parameter constraints further \cite{Hannestad:2004ts}.
\end{enumerate}

\section{Conclusion}
\label{sec:conclusions}

We believe that this analysis is the most thorough investigation of
the detectability of a trans-Planckian modulation to the primordial
power spectrum that has been performed to date.   We consider three
complementary approaches -- a simple grid search, a Fisher matrix
evaluation of the likely error ellipses, and a Monte Carlo Markov
Chain fit to a simulated CMB spectrum, and find that they are all in
broad agreement.    Moreover, we also explain the ``islands'' seen in
the likelihood space in previous papers on this topic
\cite{Elgaroy:2003gq,Okamoto:2003wk}, and show that their distribution
and properties can be understood and reproduced via a simple analytic
argument.

The approach we have taken here is analyze a {\em specific\/} ansatz
for introducing a minimum length into the calculation of the
perturbation spectrum in a general model of slow-roll inflation
\cite{Easther:2002xe}. We do not claim that this model is a correct
description of the trans-Planckian contribution to the perturbation
spectrum, but intend it as a case study of what might be possible if
one has a specific and well-motivated correction to the spectrum that
leads to a $k$ dependent modulation.   There is  considerable
theoretical uncertainty surrounding  this point, and it is entirely
possible that a future rigorous calculation of the perturbation
spectrum within string theory or some other model of ultra high energy
physics will predict that there is {\em no\/} modulation to be
observed, and the bound calculated here will not apply.

We believe that the constraint obtained here is applicable to a
general class of modulated spectra. In practice, one can imagine
adding an arbitrary modulation to the primordial spectrum, as
considered by Okamoto and Lim \cite{Okamoto:2003wk}, in which case one
has parameters which describe the amplitude, wavelength, and phase of
the modulation as a function of $k$.   In our model there are only two
free parameters -- the phase of the modulation, and $H/M$ which
determines the amplitude. The phase will always be arbitrary in the
absence of detailed theory that matches specific scales in the present
universe to those during inflation, which would require a full
understanding of the post-inflationary expansion history and the
physics of reheating, while the amplitude is directly related to the
scale of new physics, which is what we are trying to measure. However,
the wavelength of the oscillation turns out to be given in terms of
the slow-roll parameter $\epsilon$, which is itself directly related
to the tensor/scalar ratio $r$. This dependence is easy to understand,
since the form of the modulation is not determined by the (fixed)
boundary condition, but by the slow variation in the horizon size
$H^{-1}$ with time. The precise functional form of the connection
between $\epsilon$ and the modulated spectrum may vary in other
specific models. However, it is very likely that the modulation can
always be expressed in terms of $H/M$ and the scale dependence of $H$
-- or thus $\epsilon$, and perhaps the higher order slow-roll
parameters, which are also reachable through their contribution to the
scalar spectral index.  Consequently, we believe that the results we
have seen here will generalize to other modulated spectra, even though
we have reduced the number of trans-Planckian parameters relative to
those considered by Okamoto and Lim.

The second general conclusion we draw is that the detectability of any
modulated spectrum depends on the value of $H$.  Since the wavelength
of the modulation depends on $\epsilon$, detecting the primordial
tensors provides an orthogonal  constraint on the value of this
parameter, and thus constrains the trans-Planckian corrections. Since
$H$ fixes the energy scale at which inflation occurs, it is arguably
the single most interesting cosmological parameter that is not
currently fixed by observations.  The analysis here simply adds to the
importance of this parameter, since we have shown that measuring $H$
will put tighter bounds on any trans-Planckian corrections to the
spectrum.

Since the theoretical uncertainty about the form and existence of
these corrections is yet to be resolved, we cannot advocate mounting
an observational campaign solely to look for this kind of signal.
However, there is already good and sufficient reason to make an all
out effort to measure $H$ during inflation, via high precision
measurements of the CMB and particularly its polarization.   We
believe that mission planners will want to be aware of the possibility
that precision CMB measurements may potentially probe physics at the
string scale,  since this will not add to the cost of any mission  but
represents an exciting and additional use of the data   they can be
expected to return.

In quantitative terms, we see that the with a large ($r\sim 0.15$)
value of the tensor to scalar ratio, a ``perfect'' map of the
primordial temperature and polarization anisotropies  to the CMB could
rule out a trans-Planckian modulation to the spectrum at the $2
\sigma$ level for $H/M \sim 0.004$.  This is roughly the same as the
result reported by  Okamoto and Lim.  We vary a larger set of
cosmological parameters than they do, which reduces the level of
precision we can hope for. However by adding the tensor contribution
to the analysis and removing a free parameter from the specification
of modulation we can tighten the bound.   Like Okamoto and Lim, we
consider a perfect measurement of the CMB and any real experiment will be
contaminated by imperfectly removed foregrounds.   Consequently, we
believe that a value of $H/M$ on the order of 0.01 represents a
reasonable lower bound on what can be detected in practice, provided
the amplitude of the tensor contribution is significant.   On the
other hand, adding other orthogonal datasets, such as a measurement of
$h$ and the late time expansion history of the universe from a
SNAP-like mission, or constraints on the primordial spectrum from
future large scale structure surveys will further constrain the error
ellipses relative to those which can be achieved via the CMB alone.

At this point,   the theoretical uncertainty surrounding this
calculation makes it unwise to   push this analysis significantly
further. However, we believe that considerable progress has been made
in understanding the physical issues surrounding the trans-Planckian
corrections to the primordial spectrum over the last few
years. Consequently,  it is not excessively optimistic to hope that
the remaining issues can be resolved on a shorter timescale than it
will take to perform a measurement of the CMB that even approaches the
precision we have assumed in our analysis.    In this light, we are
particularly interested in the effective field theory approach taken
in \cite{Schalm:2004qk}, and extended in the more recent preprint
\cite{Greene:2004np}.  In this case, the operators which can
contribute to a modification to the spectrum can be cataloged within
effective field theory. From the observational perspective, an
analogous calculation to the one presented here will constrain  the
values of the prefactors in front of these operators, and we intend to
pursue this in future work.

\section*{ Acknowledgments}  We thank  Brian Greene, Gaurav Khanna,
Eugene Lim, Alexey Makarov, Jerome Martin, Koenraad Schalm, and Uros
Seljak for useful discussions, and are grateful to the staff at Yale's
High Performance Computing facility for their assistance. RE  is
supported in part by the United States Department of Energy, grant
DE-FG02-92ER-40704. HVP is supported by NASA through Hubble Fellowship
grant \#HF-01177.01-A awarded by the Space Telescope Science
Institute, which is operated by the Association of Universities for
Research in Astronomy, Inc., for NASA, under contract NAS 5-26555; she
acknowledges the hospitality of the Institute of Astronomy, Cambridge,
where part of this work was carried out. WHK acknowledges the hospitality
of the Perimeter Institute, Waterloo, Ontario, where part of this work
was carried out.

 \end{document}